\documentclass[
aps
,groupedaddress
,amsfonts,amsmath,floatfix,eqsecnum
,nofootinbib
,twocolumn
]{revtex4}

\usepackage{hyperref}

\usepackage[utf8]{inputenc}
\usepackage{epsfig,graphicx}
\usepackage{bm}
\usepackage{amsmath,amsfonts}
\usepackage{enumerate}
\usepackage{ytableau}
\usepackage{cancel}
\usepackage{multirow}
\usepackage{mathptmx}      
\usepackage{latexsym}
\usepackage{amstext}
\usepackage{array}
\usepackage[T1]{fontenc} 

\DeclareMathSymbol{\gtrless} {\mathrel}{AMSa}{"3F}

\newcommand{\ignore}[1]{}
\newcommand{\nec}[1]{\eqref{eq:#1}}

\newcommand{\eq}[1]{eq. \eqref{eq:#1}}

\newcommand{\Eq}[1]{Eq. \eqref{eq:#1}}
\newcommand{\Eqs}[1]{Eqs. \eqref{eq:#1}}
\newcommand{\be}{\begin{equation}}
\newcommand{\ee}{\end{equation}}
\def\bes#1\ees{%
  \begin{equation}
    \begin{split}
      #1
    \end{split}
  \end{equation}
}
\def\bs#1\es{%
    \begin{split}
      #1
    \end{split}
}

\newcommand{\R}{\mathbb{R}}

\renewcommand{\S}{\mathrm{S}}

\newcommand{\SO}{\mathrm{SO}}
\newcommand{\SU}{\mathrm{SU}}
\newcommand{\U}{\mathrm{U}}

\newcommand{\mfootnote}[1]{\footnote{\sf #1}}

\def\slashchar#1{\setbox0=\hbox{$#1$}
   \dimen0=\wd0 \setbox1=\hbox{/} \dimen1=\wd1
   \ifdim\dimen0>\dimen1 \rlap{\hbox to \dimen0{\hfil/\hfil}} #1
   \else  \rlap{\hbox to \dimen1{\hfil$#1$\hfil}} / \fi}

\newcommand{\ben}{\begin{enumerate}}
\newcommand{\een}{\end{enumerate}}

\DeclareMathOperator{\Det}{Det}
\DeclareMathOperator{\Tr}{Tr}
\DeclareMathOperator{\tr}{tr}

\DeclareMathOperator{\diag}{diag}

\newcommand{\cA}{{\mathcal A }}
\newcommand{\cB}{{\mathcal B }}
\newcommand{\cC}{{\mathcal C }}
\newcommand{\cD}{{\mathcal D }}
\newcommand{\cF}{{\mathcal F }}

\newcommand{\cH}{{\mathcal H }}

\newcommand{\cL}{{\mathcal L }}
\newcommand{\cM}{{\mathcal M }}

\newcommand{\cO}{{\mathcal O }}

\newcommand{\cR}{{\mathcal R }}

\newcommand{\bR}{{\mathbf{R}}}

\newcommand{\esp}[1]{\langle #1 \rangle}
\newcommand{\Esp}[1]{\left\langle #1 \right\rangle}

\newcommand{\gh}{\mathrm{gh}}

\renewcommand{\div}{\mathrm{div}}
\newcommand{\ang}{\mathrm{ang}}

\newcommand{\Gammadiv}{\Gamma^\div}
\renewcommand{\neq}{\nec}
\newcommand{\ZR}{Z^R}
\newcommand{\Deltag}{\Delta}
\newcommand{\DeltaM}{\Delta_M}
\newcommand{\oDeltaM}{\overline{\Delta}_M}
\newcommand{\hk}{\hat{k}}
\newcommand{\hcA}{\hat{\cA}}
\newcommand{\hcF}{\hat{\cF}}
\newcommand{\RBV}{\hat{\cR}}
\newcommand{\unit}{\mathbf{1}}

\renewcommand{\mfootnote}[1]{\footnote{\rm #1}}

\begin{document}

\title{\textsf{
Non-minimal non-Abelian quantum vector fields in curved spacetime
}}

\author{L. L. Salcedo}
\email{salcedo@ugr.es}

\affiliation{Departamento de F\'{\i}sica At\'omica, Molecular y Nuclear and \\
  Instituto Carlos I de F\'{\i}sica Te\'orica y Computacional, \\ Universidad
  de Granada, E-18071 Granada, Spain.
}

\date{\today}

\begin{abstract}
  The quantum effective action of non-minimal vector fields with Abelian or
  non-Abelian gauge degrees of freedom in curved spacetime is studied. The
  Proca or Yang-Mills fields are coupled to a local mass-like term acting in
  both coordinate and gauge spaces.  Pathologies due to gauge invariance in
  the ultraviolet are avoided through the introduction of a non-Abelian
  version of the Stueckelberg field. It is found that the breaking of gauge
  invariance induced by the mass term affects only the tree level part of the
  effective action. The ultraviolet divergent part of the effective action to
  one loop is obtained using the method of covariant symbols and dimensional
  regularization. Formulas are given valid for any spacetime dimension and
  explicit results are shown for the two-dimensional case. As already happened
  for a single vector field, the ultraviolet divergences are local but not of
  polynomial type.
\end{abstract}

\keywords{}



\maketitle
\flushbottom
\setlength{\unitlength}{1mm}

\tableofcontents

\section{\textsf{Introduction}}
\label{sec:1}

Vector fields play a prominent role in the Standard Model of particles, as
mediators of gauge interactions. In turn there is currently a growing interest
on the role that various types of vector fields could play in relativistic
gravity and cosmology \cite{Dimopoulos:2006ms, Dimopoulos:2008yv,
  Koivisto:2008xf, Yokoyama:2008xw, Esposito-Farese:2009wbc, Graham:2015rva,
  DeFelice:2016yws, Emami:2016ldl, Heisenberg:2018vsk}. As noted in
\cite{Heisenberg:2018vsk} ``Imposing the conditions of Lorentz symmetry,
unitarity, locality and a (pseudo-)Riemannian space–time, any attempt of
modifying gravity inevitably introduces new dynamical degrees of freedom. They
could be additional scalar, vector or tensor fields.'' The subject has
received a further boost with the discovery of ghost free consistent
non-linear theories of Proca interactions \cite{Heisenberg:2014rta,
  Allys:2015sht, Allys:2016kbq, ErrastiDiez:2019ttn, deRham:2020yet,
  GallegoCadavid:2020dho}. The crucial issue of the quantum stability of these
theories has been analyzed in \cite{Heisenberg:2020jtr, deRham:2021yhr}.

In this work we consider a set of $N$ vector fields in curved spacetime
endowed with Abelian (Proca) and/or non-Abelian (Yang-Mills) internal degrees
of freedom. No self-interactions are included beyond those implied by the
Yang-Mills structure, but the vector fields are coupled to an external
mass-like $x$-dependent tensor field which is allowed to arbitrarily mix them
(see \Eq{2.1}). Our focus is on the proper quantization of such a theory and
on the structure of the quantum fluctuations.

Early work studying the subject of quantum fluctuations for vectors fields was
carried out in \cite{Barvinsky:1985an, Furlani:1997fh, Gorbar:2003yt,
  Buchbinder:2007xq} (see \cite{ Felipe:2018pdw, Franchino-Vinas:2019upg,
  Gomez:2019tbj, Sanongkhun:2019ntn, ErrastiDiez:2020dux, Marques:2021jfy,
  Martini:2021slj, Hell:2021oea}
for recent related work). Particular non-minimal couplings (the minimal case
being a standard mass term) were considered in \cite{Novello:1979ik} at the
classical level and in \cite{Davies:1984vm} at the quantum level. The
quantized theory for general non-minimal couplings was first studied in
\cite{Toms:2015fja} for particular spacetime backgrounds. There it was found
that pathologies arise in the quantization of the theory since the mass term
couples effectively as a metric field.  Technically the problem is that the
mass-like field breaks gauge invariance but does not suppress the fluctuations
in the longitudinal polarization at large wavenumbers. In other words, the
principal symbol of the fluctuation operator is singular. General backgrounds
were considered in \cite{Buchbinder:2017zaa} solving the above-mentioned
pathology by means of a Stueckelberg field. In this way a proper gauge
symmetry is present in the theory and one can proceed through a standard gauge
fixing procedure. However, approximations were introduced in the analysis of
\cite{Buchbinder:2017zaa} giving rise to a non-local result. A full solution
to the problem of computing the ultraviolet (UV) divergent part of the
effective action, within dimensional regularization, was obtained in
\cite{Ruf:2018vzq} using the Schwinger-DeWitt technique and later in
\cite{Garcia-Recio:2019iia} using the method of covariant symbols, finding
perfect concordance in both calculations. The pathologies identified in
\cite{Toms:2015fja} translate to the fact that the UV divergences are local
but not polynomial in the mass-like external field.

The results just noted refer to a single vector field. Here we address the
case of several vector fields. This allows to consider the non-Abelian
scenario. In fact, we consider sets of vector fields organized in Abelian and
non Abelian multiplets. We treat the Abelian and non-Abelian versions
simultaneously since the formalism is identical in both cases. In the absence
of a mass term there would be a gauge symmetry present in the Lagrangian. This
symmetry is explicitly broken by the mass term. Nevertheless we obtain the
remarkable result that the breaking only affects to the effective action at
tree level, while the contributions from one or more loops are fully gauge
invariant. This results in an important simplification of the
calculation. Another insight comes from the introduction of the Stueckelberg
field in our present non-Abelian setting (see \cite{Ruegg:2003ps} for a review
on this subject). In the Abelian case the Stueckelberg field appears through
$A_\mu=B_\mu+\partial_\mu\phi$. In this way a $U(1)$ gauge symmetry arises
from $B_\mu\to B_\mu+\partial_\mu \Lambda$, $\phi \to \phi - \Lambda$. In the
non-Abelian case a literal translation of this prescription would take the
form $A^a_\mu = (B^\Omega)^a_\mu$ where $\Omega$ refers to a non-Abelian gauge
transformation with parameters $\phi^a$ (in Lie algebra of the gauge
group). The resulting theory enjoys a non-Abelian gauge invariance and one can
then proceed to fix the gauge through the Fadeev-Popov method. Such approach
is in principle correct but exceedingly complicated as the dependence on
$\phi^a$ is non-linear. In particular this would imply a reorganization of the
loop expansion from the original theory (field $A_\mu^a$) to that with
$B_\mu^a$ and $\phi^a$. We develop a completely different approach where the
Stueckelberg field is introduced linearly also in the non-Abelian setting.

Once the Stueckelberg field is introduced the (UV regulated) quantum theory is
no longer pathological and it is possible to proceed to a systematic
computation of its effective action. Our focus is on the UV part of the
effective action to one-loop order.  As already known from the study of the
Abelian case \cite{Ruf:2018vzq,Garcia-Recio:2019iia}, the mass term acts as an
effective second metric tensor. In the present case this is in fact a
non-Abelian effective metric. Presumably the generalized Schwinger-DeWitt
technique \cite{Barvinsky:1985an} can be adapted to this situation, but such an
approach is not presently available. Instead here we apply the method of
covariant symbols \cite{Salcedo:2006pv}. This method is simple to use and
allows to formulate the loop momentum integration while preserving manifest
covariance under diffeomorphism and gauge transformations. Details of the
method are provided below.

The problem studied here is an extension of that already solved for $N=1$
(just one vector field), so some specific features of that case are inherited
in the more general setting analyzed in this work. In particular, the loop
momentum integrals cannot be written in closed form. For $N>1$ this problem is
even worse as the propagator are now matrices with respect to the gauge
indices. Also for this reason some contributions to the effective action (see
$\Gamma_{L,0}$ below) cannot be expressed in a standard form involving just
integration over $x$, $p$ and traces in internal space, and it is necessary to
resort to a parametric form, with integration over one more
parameter. Unfortunately, while the problem is well-posed and the method fully
appropriate to solve it, we have found an unexpected impediment, namely, the
number of terms obtained for the physically relevant case of four spacetime
dimensions is prohibitively large (at least hundreds of terms are
generated). In view of this we develop the formulas for the general case but
only present detailed results for two spacetime dimensions (note that there
are no UV divergences for odd dimensions within dimensional regularization).

In Sec. \ref{sec:2.1} we expose the theory to be analyzed.
In Sec. \ref{sec:2.2} the background field approach is introduced for the
effective action. It is shown that its quantum part admits a gauge-covariant
treatment.  The effective action to one loop is constructed, showing its
limitations in the UV sector.
Those obstacles are overcome in Sec. \ref{sec:2.3} by introducing the
Stueckelberg field. The non-pathological one-loop effective action is
constructed, and then decomposed into various contributions to be computed
subsequently.
Sec. \ref{sec:31} introduces some notational conventions.
Sec. \ref{sec:9} introduces general considerations to undertake the calculation
and presents explicit results for $d=2$. Some non-trivial symmetries related
to metric deformations are also verified.
The actual calculations are worked out in
Sec. \ref{sec:12}. To this end, the method of covariant symbols is
reviewed first, and its application to the various contributions
is discussed, including the extraction of the coefficients of the UV divergence.
The conclusions are summarized in Sec. \ref{sec:sum}.
The proof of some formulas is provided in App. \ref{app:b}. Properties of the
operator $Z_{\mu\nu}$ and its relation to $\hat{\cR}_{\mu\nu}$ of
\cite{Barvinsky:1985an} are discussed in App. \ref{app:a}. The canonical form
of $\Gamma_S$ using a basis of standard operators is displayed in
App. \ref{app:c}. Explicit results for perturbative mass expansions are
presented in App. \ref{app:d}. Details of the method of non-covariant symbols,
used as a check of the calculations, are given in App. \ref{app:e}.  Finally,
the method of covariant symbols is illustrated through a sample computation in
App. \ref{app:f}.

\section{\textsf{Formulation of the problem}}
\label{sec:2.1}

We consider $N$ real vector fields $\hcA^a_\mu(x)$, $a=1,\ldots,N$\/ in an
Euclidean $d$ dimensional spacetime with metric $g_{\mu\nu}(x)$ and
action\mfootnote{The ugly notation $\hcA^a_\mu$ and $\hcF_a^{\mu\nu}$ will soon
  be traded by $A^a_\mu$ and $F^{\mu\nu}$.}
\be
S[\hcA;\cM,g] = \int d^d x \sqrt{g}\left(
\frac{1}{4}\hcF_a^{\mu\nu}\hcF^a_{\mu\nu}
+ \frac{1}{2} \cM_{ab}^{\mu\nu} \hcA^a_\mu \hcA^b_\nu
\right)
,
\label{eq:2.1}
\ee
where $\cM_{ab}^{\mu\nu}(x)$ is a positive definite\mfootnote{As a matrix with
  indices $(\mu a)$ and $(\nu b)$.} local mass term fulfilling
the symmetry condition
\be
\cM_{ab}^{\mu\nu}(x) = \cM_{ba}^{\nu\mu}(x)
.
\ee
Coordinate indices are raised, lowered and contracted with the metric
$g_{\mu\nu}$.

The $N$ fields are organized in $n$ gauge sectors. Each sector has gauge
symmetry of the type either $\SU(n_i)$ or $\U(1)$ and the full gauge group is
the direct product of these.\mfootnote{More generally, one could take a Lie
  subgroup of $\SO(N)$ and the results and formulas derived in this work hold
  equally well in that case.} The fields fall in the Lie algebra of the group,
i.e., the adjoint representation, with gauge coupling $g_i$ in the gauge
sector $i$. Without loss of generality one can choose $g_i=1$ for the Abelian
factors. Hence $N=\sum_{i=1}^n N_i$ where $N_i=n_i^2-1$ for an $\SU(n_i)$
sector and $N_i=1$ for a $\U(1)$ one. With a standard normalization of the
fields, the field strength tensor is
\be
\hcF^a_{\mu\nu} = \partial_\mu \hcA^a_\nu - \partial_\nu \hcA^a_\mu
+ g_a f_{abc} \hcA^b_\mu \hcA^c_\nu
\label{eq:2.3}
\ee
where $f_{abc}$ are the structure constants of the gauge group. Since the
gauge group is a direct product, the structure constants are block diagonal,
one block for each gauge sector, and $g_a=g_i$ is the coupling of the $i$-th
sector (the $g_a$ take a common value within each block). Of course, in a
$\U(1)$ sector the structure constant vanishes. The field strength tensor
$\hcF^a_{\mu\nu}(x)$ is covariant under gauge and coordinate transformations.
The kinetic part of the action is block diagonal while the mass term may mix
different gauge sectors.

When all the gauge sectors are of the $\U(1)$ type the theory is Abelian and
reduces to a generalized Proca field with $N$ flavors. Nevertheless, since all
the cases can be treated within the same scheme, we will refer to the internal
space as gauge space.

Regarding the symmetries, the kinetic term is fully local gauge invariant but
such symmetry is reduced to a global one by the mass term: the action is
invariant under $\cM\to \cM^\Omega = \Omega \cM \Omega^{-1}$ for a global
transformation $\Omega$ in the gauge group. If some gauge sectors are
equivalent, namely, with equal gauge group $\SU(n_i)$ o $\U(1)$ and same
$g_i$, there is an additional global symmetry under rotations among those
equivalent sectors, with a corresponding rotation of $\cM$.  A further
symmetry, special for the case of $d=4$ spacetime dimensions, is that of local
Weyl-like transformations, namely, the action is unchanged under the
simultaneous replacements $g_{\mu\nu}(x) \to \xi(x) g_{\mu\nu}(x)$ and
$\cM^{\mu\nu}_{ab}(x)\to \xi^{-2}(x)\cM^{\mu\nu}_{ab}(x)$.

\section{\textsf{The effective action }}
\label{sec:2.2}

\subsection{\textsf{The background gauge field}}

Within the background field approach \cite{Abbott:1980hw} the field is split
as a background plus a fluctuation, $\hcA^a_\mu(x) = A^a_\mu(x) +
\cA^a_\mu(x)$.

In this approach the effective action $\Gamma[A;\cM,g]$ follows from
\be
Z = e^{-\Gamma[A;\cM,g]} = \int \cD \cA e^{-S[\cA+A]+\int d^d x \sqrt{g}\,J\cA}
\label{eq:3.1}
\ee
where $A_\mu^a(x)$ is the background field and the current $J^\mu_a(x)$ is
adjusted so that $\esp{\cA_\mu^a(x)}=0$. As usual
\be
J^\mu_a(x) = \frac{1}{\sqrt{g(x)}} \frac{\delta \Gamma[A;\cM,g]}{\delta
  A_\mu^a(x)}
.
\ee

In the background gauge field approach, the field $\cA_\mu^a(x)$ transforms
homogeneously under local gauge transformations, the inhomogeneity being
saturated by the transformation of $A^a_\mu(x)$. Correspondingly, the gauge
covariant derivative relies on $A^a_\mu(x)$ as gauge connection.

We will use a single covariant derivative $\nabla_\mu$ containing coordinate
and gauge connections \cite{Vassilevich:2003xt}. The coordinate connection is
that of Levi-Civita for the metric and the gauge connection is that of the
background field $A_\mu^a$. So for instance, for coordinate-scalar and
coordinate-vector fields $\phi^a$ and $\cB^a_\mu$, respectively, both in the
adjoint gauge representation,
%
\bes
\nabla_\mu \phi^a
&
= \partial_\mu \phi^a + g_a f_{abc} A_\mu^b \phi^c
,
\\
\nabla_\mu \cB_\nu^a
&
= \partial_\mu \cB_\nu^a
- \Gamma_{\mu\nu}^\lambda \cB_\lambda^a
+ g_a f_{abc} A_\mu^b \cB_\nu^c
.
\ees
Throughout, coordinate indices are contracted with the metric $g_{\mu\nu}$ and
gauge-vector indices with $\delta_{ab}$.

The effective action can be split into the classical or tree-level component
$S[A]$ and the quantum correction $\Gamma_Q[A]$ which contains graphs with one
or more loops,
\be
\Gamma[A] = S[A] + \Gamma_Q[A]
.
\ee
Here we find a fundamental result given by the following

{\bf Theorem:} ~ $\Gamma_Q[A]$ is invariant under {\em local} gauge
transformations and all the gauge breaking in the effective action is
saturated by the mass term at tree level. That is,
\be
\Gamma_Q[A;\cM,g] = \Gamma_Q[A^\Omega;\cM^\Omega,g]
,
\label{eq:3.5}
\ee
where $\Omega(x)$ is any local gauge transformation, and $A^\Omega$,
$\cM^\Omega$ are the gauge-transformed fields.

{\bf Proof:} The reason is fairly
simple. The semiclassical expansion follows from a Taylor expansion of the
action $S[\cA+A]-\int d^d x \sqrt{g}\,J\cA$ in powers of the fluctuation
$\cA$. The zeroth order gives the classical action, and the first order in
$\cA$ cancels due to the equations of motion, i.e. the choice of $J^\mu$.  The
quantum component $\Gamma_Q$ depends only on terms which are quadratic or
higher order in $\cA$. The breaking of gauge invariance would come solely from
the mass term $\frac{1}{2}\cM_{ab}^{\mu\nu} \cA^a_\mu\cA^b_\nu$ but this is
covariant since the field $\cA_\mu^a$ transforms homogeneously under gauge
transformations.

The property \Eq{3.5} is important because it allows to use a gauge-covariant
formalism for $\Gamma_Q$.

\subsection{\textsf{One-loop effective action}}
\label{sec:2.4}

The one-loop effective action follows from the quadratic part of the action
\be
\Gamma_1[A] = \frac{1}{2}\log\Det_1\left(\frac{\delta^2 S[\cA+A]}{\delta
    \cA{}^2}\right)\Bigg|_{\cA=0}
\ee
The subindex $1$ in $\Det_1$ indicates that the determinant is to be evaluated
in the space $\cA_\mu(x)$, i.e., of coordinate vectors. Also, for the gauge
degrees of freedom, the determinant is taken in the adjoint gauge
representation space. This is not explicitly indicated but will be implicit in
all formulas as no other gauge representations will be present.

Therefore we need to isolate the terms quadratic in $\cA$ from the action
$S[\cA+A]$. After the shift $\hcA = \cA+A$, the field strength tensor in
\Eq{2.3} becomes
\be
\hcF^a_{\mu\nu} =
\cF^a_{\mu\nu} + \nabla_\mu\cA^a_\nu - \nabla_\nu \cA^a_\mu
+ g_a f_{abc} \cA^b_\mu \cA^c_\nu
\ee
with
\be
\cF^a_{\mu\nu} = \partial_\mu A^a_\nu - \partial_\nu A^a_\mu
+ g_a f_{abc} A^b_\mu A^c_\nu
.
\ee

In the shifted variables, the quadratic part of the kinetic term of the action
takes the form\mfootnote{We will occasionally place all the coordinate indices
  as lower indices when no ambiguity arises. Repeated coordinate indices are
  always contracted with the metric $g_{\mu\nu}$.}
\be
S^{(2)}_{\rm kin}[\cA] = \int d^d x\sqrt{g} \left(
 \frac{1}{4} (\nabla_\mu\cA_\nu^a-\nabla_\nu\cA_\mu^a)^2
-\frac{1}{2}  F_{\mu\nu}^{ab} \cA_\mu^a \cA^b_\nu
\right)
\ee
where we have the introduced field strength tensor $F^{ab}_{\mu\nu}$ as an
antisymmetric matrix in gauge space (as well as in coordinate space)
\be
F_{\mu\nu}^{ab} \equiv g_c f_{acb} \cF^c_{\mu\nu}
.
\ee
Note that $F_{\mu\nu}^{ab}$ vanishes in the Abelian sectors.

In what follows we adopt the convention that covariant derivatives are
indicated by adding new indices {\em to the left}, hence
$\phi_\mu^a \equiv \nabla_\mu \phi^a$,
$\cB_{\mu\nu}^a \equiv \nabla_\mu \cB^a_\nu$, etc. The only exception to this
rule are the operators $Z_{\mu_1\mu_2\ldots}$ and
$Z^R_{\mu_1\mu_2\ldots}$.\mfootnote{Some conventions used in this work are
  summarized in Sec. \ref{sec:31}.} With this convention
\be
S^{(2)}_{\rm kin}[\cA] = \int d^d x\sqrt{g} \left(
 \frac{1}{4} (\cA_{\mu\nu}^a-\cA_{\nu\mu}^a)^2
-\frac{1}{2}  F_{\mu\nu}^{ab} \cA_\mu^a \cA^b_\nu
\right)
\label{eq:2.20a}
\ee

Using integration by parts and Bianchi identities, the quadratic part of the
kinetic term can be written as (see App. \ref{app:b})
%
\bes
S^{(2)}_{\rm kin}[\cA]
&
= \int d^d x\sqrt{g}
\Bigg(%
 \frac{1}{2}(\cA_{\mu\nu}^a)^2
- \frac{1}{2}(\cA_{\mu\mu}^a)^2
\\&\quad%
- F_{\mu\nu}^{ab} \cA_\mu^a \cA^b_\nu
+ \frac{1}{2} \cR_{\mu\nu}\cA^a_\mu\cA^a_\nu
\Bigg)%
\label{eq:2.20}
\ees
where $\cR_{\mu\nu}$ is the Ricci tensor.  Thus, adding the mass term,
\be
S^{(2)}_{\rm mass}[\cA] = \int d^d x\sqrt{g} \left(
\frac{1}{2} \cM_{ab}^{\mu\nu} \cA^a_\mu \cA^b_\nu
\right)
,
\ee
the full quadratic Lagrangian controlling the one-loop fluctuations is
\be
\cL^{(2)}(x) = \frac{1}{2} \cA_\mu K^{\mu\nu}_0 \cA_\nu
\ee
where
\be
K_0^{\mu\nu} = - g^{\mu\nu} \nabla^2 +\nabla^\mu\nabla^\nu
- 2 F^{\mu\nu} + \cR^{\mu\nu} + \cM^{\mu\nu} 
.\ee
Here, and also in what follows, we use a matric notation for the gauge
indices, which will be implicit. As advertised, the Lagrangian $\cL^{(2)}(x)$
is manifestly gauge invariant.

The term $+\nabla^\mu\nabla^\nu$ in $K_0^{\mu\nu}$ is a direct consequence of
gauge invariance of the kinetic energy part of the action \nec{2.1} and is
needed to retain just three polarizations in the Proca field. While the
differential operator $K_0$ needs not be singular in presence of a positive
definite mass term $\cM^{\mu\nu}$, its principal symbol, i.e., the
$O(\nabla^2)$ leading UV divergent component is singular, since the
longitudinal polarizations are not penalized at large wavenumbers. The fact
that the principal symbol is singular introduces pathologies in the effective
action which prevent to carry out an extraction of the UV divergent terms.

In the special case of a standard Proca field, with a constant scalar mass,
the UV divergent part of the effective action is a polynomial in the mass
\cite{Barvinsky:1985an}, but this is no longer so for a non-constant mass term
even in the Abelian case \cite{Ruf:2018vzq}. This confirms that $K_0$ cannot
be directly used as the fluctuation operator.

\section{\textsf{The Stueckelberg field}}
\label{sec:2.3}

\subsection{\textsf{The non-Abelian Stueckelberg field}}

In order to bypass the above mentioned pathologies in the UV we will adapt the
Stueckelberg approach introduced in \cite{Buchbinder:2017zaa} (and also
applied in \cite{Ruf:2018vzq,Garcia-Recio:2019iia}) for the
non-minimal Proca field to the non-Abelian case. To this end, we rewrite the
partition function as
\be
Z = \int \cD \cA \, e^{-S[\cA+A] + \int d^d x
  \sqrt{g}\,J\cA}
\int \cD\chi \, e^{-S_{\rm gf}[\chi]}
\label{eq:5.1}
\ee
where $S_{\rm gf}[\chi]$ can be any action. The partition function is
unchanged as the new factor is just a constant.\mfootnote{It does not depend
  on $\cA_a^\mu(x)$, $J^\mu_a(x)$ nor $\cM^{\mu\nu}_{ab}(x)$. It is also
  independent of the metric if no derivatives are involved.} We take a
standard choice
\be
S_{\rm gf}[\chi] = \int d^d x \sqrt{g} \frac{1}{2} \chi^a\chi_a
,
\ee
where $\chi^a(x)$ is a coordinate scalar and a gauge vector (i.e., in the
adjoint representation).

Subsequently a change of variables $(\cA,\chi) \to (\cB,\phi)$ is applied in
\nec{5.1}, where $\cB_\mu^a(x)$ is a real coordinate-vector and gauge-vector
field and $\phi^a(x)$ is real coordinate-scalar and gauge-vector field:
%
\bes%
&%
Z
= \int \cD \cB \,\cD\phi \, 
J[\cB,\phi]
\,
e^{-S[\cA+A]-S_{\rm gf}[\chi] + \int d^d x \sqrt{g}\,J\cA},
\\%
&%
J[\cB,\phi]
= \Det\left(\frac{\partial(\cA,\chi)}{\partial(\cB,\phi)} \right)
.
\ees
By construction the effective action does not depend on the detailed choice of
gauge-fixing function(al) $\chi[\cB,\phi]$, moreover, the expectation value of
any functional written in the form $F[\cA,\chi]$ is independent of this choice
(unless the very functional $F$ depends on it). This property provides
identities for the gauge-fixing dependence of the expectation values
\cite{Collins:1984xc}.

We choose a linear change of variables. Besides simplicity, the virtue of such
a choice is that the loop expansion in the new variables coincides with that
in the old ones. Specifically we take
\be
\cA_\mu^a = \cB_\mu^a + \nabla_\mu \phi^a,
\qquad
\chi^a = \nabla_\mu \cB^a_\mu
,
\label{eq:5.4}
\ee
or, using our notational convention for the covariant derivatives,
\be
\cA_\mu^a = \cB_\mu^a + \phi_\mu^a,
\qquad
\chi^a = \cB^a_{\mu\mu}
.
\label{eq:2.13}
\ee

The corresponding Fadeev-Popov determinant is easily obtained
as (see App. \ref{app:b})
\be
J[\cB,\phi] = \Det_0 \left( \delta_{ab} \,
\nabla^\alpha \nabla_\alpha \right).
\label{eq:2.14}
\ee
The subindex $0$ in $\Det_0$ indicates that the determinant is to be evaluated
in the $\phi^a$ space, i.e. the coordinate-scalar space. As already noted, the
reference to the adjoint gauge representation is not explicitly displayed as
its presence is ubiquitous and no other gauge representation will be needed.
With our choice of a linear change of variables, the determinant $J$ does not
depend on the quantum fields $\cB,\phi$. It depends on $A_\mu^a$ and
$g_{\mu\nu}$. As usual the determinant can be implemented through a complex
ghost field with quadratic action.

It is worth noticing that one could have introduced the Stueckelberg field in
a different manner, to wit, through the change of variable $\cA=\cB^\Omega$ in
\Eq{3.1}, where $\Omega(x)$ is an arbitrary gauge transformation, and
$\phi^a(x)$ enters through $\Omega=e^{i\phi}$. In addition the measure
$\cD\cA$ is replaced by $\cD\cB\cD\Omega$.\mfootnote{Or just
  $\cD\cB\cD\phi$. The two measures $\cD\Omega$ and $\cD\phi$ are
  equivalent. As is well-known the Jacobian of a ultra-local change of
  variables such as $\partial\Omega/\partial\phi$ has no effect in dimensional
  regularization.}  In this way the full theory $S[\cB^\Omega]$ becomes gauge
invariant even in presence of the mass term. Then one fixes the gauge as usual
with the Fadeev-Popov method. A more involved question is how to introduce the
background gauge machinery. In such alternative approach the change of
variables from $(\cA,\chi)$ to $(\cB,\phi)$ is not linear and so it should be
considerably more complicated than the method adopted above. In the Abelian
case the two approaches are equivalent.

\subsection{\textsf{The one-loop effective action revisited}}
\label{sec:2.5}

The introduction of the gauge-fixing action $S_{\rm gf}[\chi]$ adds an
irrelevant constant to the effective action. Hence in variables $(\cB,\phi)$
the effective action is just
%
\bes
\Gamma_1[A]
&%
= \frac{1}{2}\log\Det_{1+0}
\left(\frac{\delta^2( S[\cA+A]+S_{\rm gf}[\chi])}{\delta
  (\cB,\phi){}^2}\right)\Bigg|_{\substack{\cB=0 \\ \phi=0}}
\\&\quad
- \log\Det_0( \delta_{ab} \,  \nabla^2)
,
\ees
where the last term comes from the Fadeev-Popov determinant. The subindex
$1+0$ indicates the direct sum of coordinate-vector and -scalar spaces.

The kinetic energy term \Eq{2.20} in variables $(\cB,\phi)$ becomes  (see
App. \ref{app:b})
%
\bes%
S^{(2)}_{\rm kin}[\cA]
&%
= \int d^d x\sqrt{g} \Bigg(
 \frac{1}{2}(\cB_{\mu\nu}^a)^2
- \frac{1}{2}(\cB_{\mu\mu}^a)^2
- F_{\mu\nu}^{ab} \cB_\mu^a \cB^b_\nu
\\&\quad%
+ \frac{1}{2} \cR_{\mu\nu}\cB^a_\mu\cB^a_\nu
+ F^{ab}_{\mu\mu\nu} \phi^a \cB^b_\nu
+ \frac{1}{2}F^{ab}_{\mu\mu\nu} \phi^a \phi^b_\nu
\Bigg)
,
\label{eq:2.22}
\ees%
%
while $S_{\rm gf}[\chi]$ is already quadratic, namely,
\be
\S_{\rm gf}[\chi] = \int d^d x\sqrt{g} \frac{1}{2}(\cB_{\mu\mu}^a)^2
.
\ee
This contribution removes the problematic longitudinal term in the kinetic
energy.  The price to pay is the introduction of a kinetic term for $\phi$
which has a metric-like coupling to the mass tensor, namely, the last term in
%
\bes
\S^{(2)}_{\rm mass}[\cA]
&%
=
\int d^d x\sqrt{g}
\Bigg(%
\frac{1}{2}\cM^{\mu\nu}_{ab}\cB_\mu^a\cB_\nu^b
+
\cM^{\mu\nu}_{ab}\phi_\mu^a\cB_\nu^b
\\&\quad%
+ \frac{1}{2}\cM^{\mu\nu}_{ab}\phi_\mu^a\phi_\nu^b
\Bigg)%
.
\ees
%

In summary, the new full quadratic Lagrangian controlling the one-loop
fluctuations is
\be
\cL^{(2)} = \frac{1}{2} (\cB,\phi) K (\cB,\phi)^T
+ \omega^*_a \nabla^2 \omega^a
.
\label{eq:6.5}
\ee
The ghost field $\omega_a$ is a complex fermionic coordinate-scalar and
gauge-vector field. On the other hand $K$ is a second order differential
operator acting on the space $(\cB,\phi)$,
%
\bes
&
K=
\\&
\begin{pmatrix}
  -\nabla^2 g^{\mu\nu}   - 2 F^{\mu\nu} + \cR^{\mu\nu} + \cM^{\mu\nu} &
  -F_{\alpha\alpha}{}^\mu + \cM^{\mu\alpha}\nabla_\alpha
  \\
  F_{\alpha\alpha}{}^\nu - \nabla_\alpha \cM^{\alpha\nu} &
  \frac{1}{2} \{ F_{\alpha\alpha}{}^\beta, \nabla_\beta \}
  - \nabla_\alpha \cM^{\alpha\beta}\nabla_\beta  
\end{pmatrix}
.
\label{eq:2.26}
\ees
%
The matric gauge indices are implicit.  $\{\;,\;\}$ denotes anticommutator.

Note that several differential operators can be read from the last term in
\Eq{2.22}, namely, $F_{\alpha\alpha}{}^\beta \nabla_\beta$, $\nabla_\beta
F_{\alpha\alpha}{}^\beta $ or $\frac{1}{2} \{F_{\alpha\alpha}{}^\beta,
\nabla_\beta\}$. All of them are
equivalent in the $\phi$-$\phi$ sector, since the matrix $F_{\mu\nu}$ is
antisymmetric and the Hilbert space spanned by $\phi^a$ is real. However the
functional integral over $\cB$ and $\phi$ is only related to the determinant
of the {\em symmetric} version of $K$, the one presented in \Eq{2.26}.

As expected, after the introduction of the Stueckelberg field, the principal
symbol of the operator $K$ is no longer singular. Nevertheless, even though
the technical problems have been sorted, the pathologies still will reflect on
the effective action, in particular one finds that the UV divergences do not
depend polynomially on $\cM$, as already happened in the case $N=1$ studied in
\cite{Ruf:2018vzq,Garcia-Recio:2019iia}.

The leading UV divergent terms, with two derivatives, are at the diagonal of
the matrix $K$. Particularly problematic will be the term $-\nabla_\alpha
\cM^{\alpha\beta}\nabla_\beta$ in the $\phi$-$\phi$ sector. Because the
leading divergence in the covariant derivatives is Abelian, only the symmetric
component of $\cM^{\mu\nu}$ is truly of second order. Hence we will introduce
the separation of the mass tensor into symmetric and antisymmetric components:
%
\bes
\cM^{\mu\nu}_{ab}
&%
= M^{\mu\nu}_{ab} + Q^{\mu\nu}_{ab},
\qquad
M^{\mu\nu}_{ab} = M^{\nu\mu}_{ab}= M^{\mu\nu}_{ba}
,
\qquad
\\%
Q^{\mu\nu}_{ab}
&%
= -Q^{\nu\mu}_{ab}= -Q^{\mu\nu}_{ba}
.
\ees
It can be noted that $M^{\mu\nu}$ must be positive definite and should
dominate $Q^{\mu\nu}$, which is not. The mass term from $Q$ is subdivergent
since it is of first order in the derivatives. Indeed, after integration by
parts,
%
\bes%
&%
\int d^d x\sqrt{g} 
 \frac{1}{2}Q^{\mu\nu}_{ab}\phi_\mu^a\phi_\nu^b
 =
 \\&\quad%
 \int d^d x\sqrt{g} \left(
 -\frac{1}{2}Q^{\mu\mu\nu}_{ab}\phi^a\phi_\nu^b
  - \frac{1}{4}Q^{\mu\nu}_{ac}F^{cb}_{\mu\nu}\phi^a\phi^b
  \right)
  .
\ees%
Then the (symmetric) fluctuation operator $K$ takes the final form
\be
K = \begin{pmatrix}
  -\nabla^2 g^{\mu\nu} + Y^{\mu\nu}
  &
  - \Phi^\mu + \cM^{\mu\alpha}\nabla_\alpha
  \\
  \Phi^\nu - \nabla_\alpha \cM^{\alpha\nu}
  &
  - \nabla_M^2
  + \frac{1}{2}\{ P^\beta ,\nabla_\beta \}
  + W
\end{pmatrix}
,
\label{eq:4.15}
\ee
where we have introduced the following shorthand notation
\bes
Y_{\mu\nu} &= \cM_{\mu\nu} - 2 F_{\mu\nu} + \cR_{\mu\nu} ,
\qquad
\Phi_\mu  = F_{\alpha\alpha\mu}
,\qquad
\\
P_\mu &= F_{\alpha\alpha\mu} - Q_{\alpha\alpha\mu}
,
\qquad%
W = - \frac{1}{4} \{ Q_{\mu\nu} , F_{\mu\nu} \} 
,
\qquad
\ees
as well as
\be
\nabla_M^2 \equiv \nabla_\alpha M^{\alpha\beta}\nabla_\beta
.
\ee

As already noted, the field $M^{\mu\nu}(x)$, which was seemingly
UV-subdominant in the original action, is in fact UV-dominant in the sector of
the field $\phi$ in $K$ and acts effectively as a second (inverse) metric. In
the Abelian case ($N=1$ ) such metric is an ordinary one, and even so it
introduced a considerable amount of complication in the calculation of the
effective action in \cite{Ruf:2018vzq,Garcia-Recio:2019iia}. In the setting
discussed in this work the ``effective metric'' $M^{\mu\nu}(x)$ is a
non-Abelian one in gauge space, so we can certainly expect a higher degree of
difficulty in the resources needed to attack this problem.

From the Lagrangian in \eq{6.5}, the effective action to one-loop is thus
\be
\Gamma_1[A;\cM,g] =
\Gamma_K[A;\cM,g]  + \Gamma_\gh[A;g]
\ee
with
%
\bes
\Gamma_K[A;\cM,g]
&%
= \frac{1}{2}\Tr_{1+0}\log(K)
,
\qquad
\\%
\Gamma_\gh[A;g]
&
=  - \Tr_0\log(\nabla^2)
.
\label{eq:6.8}
\ees

\subsection{\textsf{ Contributions to the effective action }}
\label{sec:6}

As just said the effective action can be split into
\be
\Gamma_1= \Gamma_K  + \Gamma_\gh
.
\ee

The operator $K$ can be split into UV leading $O(\nabla^2)$ and subdivergent
$O(\nabla)$ components
%
\bes
K
&%
= K_L + K_S,
\qquad
\\%
K_L
&%
\equiv \diag (-\nabla^2 g^{\mu\nu} , - \nabla^2_M )
.
\label{eq:7.6}
\ees
Correspondingly we also separate the effective action as
\be
\Gamma_K = \frac{1}{2}\Tr_{1+0}(K_L(1+K_L^{-1}K_S)) = \Gamma_L + \Gamma_S
\ee
with
%
\bes%
\Gamma_L
&%
= \frac{1}{2}\Tr_{1+0}\log(K_L)
,
\qquad
\\%
\Gamma_S
&%
= \frac{1}{2}\Tr_{1+0}\log(1 + K_L^{-1}K_S )
.
\ees%
The relation $\Tr(\log(AB))=\Tr(\log(A))+\Tr(\log(B))$ is only
  guaranteed for sufficiently convergent pseudo-differential operators $A$ and
  $B$. Nevertheless, it is expected to correctly reproduce the UV divergent
  terms within dimensional regularization when $A$ and $B$ are both
  coordinate-scalar operators.

$\Gamma_L$ can be split further as
\be
\Gamma_L = \Gamma_{L,1} + \Gamma_{L,0}
\ee
with
%
\bes
\Gamma_{L,1}
&%
= \frac{1}{2}\Tr_1\log(-\nabla^2 g^{\mu\nu})
, \qquad
\\%
\Gamma_{L,0}
&%
=
\frac{1}{2}\Tr_0\log(-\nabla^2_M )
.
\label{eq:7.7}
\ees%

For the purpose of obtaining the UV divergences of the effective action the
subdivergent terms can be treated perturbatively
\be
\Gamma_S = \frac{1}{2}\Tr_{1+0}\log(1 + K_L^{-1}K_S )
=
\sum_{n=1}^\infty \Gamma_{S,n}
\label{eq:7.8a}
\ee
where
\be
\Gamma_{S,n} = 
-\frac{1}{2n}\Tr_{1+0}\left( (-K_L^{-1}K_S)^n  \right)
.
\label{eq:7.8}
\ee
Since $K_L^{-1}K_S=O(\nabla^{-1})$, terms with $n>d$ are UV finite in
$d$ spacetime dimensions. Thus, collecting the various contributions,
\be
\Gamma^\div_1 = \Gamma^\div_\gh + \Gamma^\div_{L,0} + \Gamma^\div_{L,1}
+ \sum_{n=1}^d \Gamma^\div_{S,n}
\,.
\label{eq:4.28}
\ee

The contributions to $\Gamma^\div_1[A;\cM,g]$ are analyzed in the following
sections, after introducing some notation.

\section{\textsf{ Some notational conventions }
\label{sec:31}}

\subsection{\textsf{ Covariant derivatives }
\label{sec:31a}}

Let us first recall that the covariant derivative operator $\nabla_\mu$ contains
all connections (and not only the Christoffel symbols), and also our
convention that covariant derivatives are indicated by adding coordinate
indices to the left. E.g.,
\be
R_{\rho\mu\nu\alpha\beta} \equiv [\nabla_\rho, R_{\mu\nu\alpha\beta} ],
\qquad
\cR_{\lambda\lambda\mu} = \frac{1}{2}\bR_\mu
.
\ee
Here $R_{\mu\nu\alpha\beta}$, $\cR_{\mu\nu}$, and $\bR$ denote the Riemann
tensor, the Ricci tensor and the scalar curvature, respectively.

All quantities in the fluctuation operator $K$ are to be regarded as operators
acting on the vector space spanned by the fields $\cB_\mu$ and $\phi$.  Hence
$\nabla_\mu$ acts on such quantities through the commutator. In particular
$g_{\mu\nu}$, $\cM^{\mu\nu}$ and $F_{\mu\nu}$ are purely multiplicative
operators, which means that they are ordinary functions (possibly matrices in
gauge space).\mfootnote{Of class $\cC(\underline{\nabla},\underline{Z})$ in
  the notation of \cite{Salcedo:2006pv}.}

\subsection{\textsf{ Operators $Z_{\mu_1\cdots\mu_n}$ }
\label{sec:31b}}

We will make use of the operator $Z_{\mu\nu}$, which is defined as
\be
Z_{\mu\nu} \equiv [\nabla_\mu, \nabla_\nu ]
.
\ee
This operator is multiplicative because its action on a quantity does not
involve derivatives of that quantity; $Z_{\mu\nu}$ is diagonal in
$x$-space. However, it is not purely multiplicative because it acts (is not
diagonal) on coordinate indices. For instance, for a purely multiplicative
tensor field $V_{\mu\nu}$
\be
[Z_{\mu\nu} ,V_{\alpha\beta}] = R_{\mu\nu\alpha\lambda}V_{\lambda\beta} +
R_{\mu\nu\beta\lambda}V_{\alpha\lambda} +[F_{\mu\nu} ,V_{\alpha\beta} ]
.
\label{eq:8.3a}
\ee
The operator $Z_{\mu\nu}$ admits a natural separation between coordinate and
gauge actions
\be
Z_{\mu\nu} = Z^R_{\mu\nu} + F_{\mu\nu}
\ee
where $Z^R_{\mu\nu}$ acts only on coordinate indices. As illustrated in
\nec{8.3a} the operator $Z^R_{\mu\nu}$ acts on every coordinate index in turn.

Higher order operators $Z_{\mu_1\cdots\mu_n}$, with $n$ covariant derivatives,
are defined recursively (see App. \ref{app:a}) so that they are also
multiplicative. Letting $I=\mu_1\cdots\mu_n$ denote a string of coordinate
indices, the operators $Z_I$ have again a clean separation between coordinate
and gauge
\be
Z_I = Z^R_I + F_I
,
\ee
and also fulfill
\be
   [Z^R_I,V_{\mu_1\mu_2}] =
   R_{I\mu_1\lambda}V_{\lambda\mu_2}
   + R_{I\mu_2\lambda}V_{\mu_1\lambda}
   .
\ee
In fact these operators are an antihermitian version of the derivatives of the
operator $\RBV_{\mu\nu}$ of \cite{Barvinsky:1985an}. Specifically
\be
Z^R_{\alpha_1\cdots\alpha_n\mu\nu} =
\nabla_{\alpha_1}\cdots \nabla_{\alpha_n}\RBV_{\mu\nu}
+
C_{\alpha_1\cdots\alpha_n\mu\nu}
,
\ee
where the $C_I$ are purely multiplicative operators constructed with the
Riemann tensor:
\bes%
C_{\mu\nu}
&%
=0,
\qquad
\\%
C_{\alpha_1\cdots\alpha_n\mu\nu}
&%
=
\frac{1}{2} R_{\lambda \alpha_2\alpha_3\cdots\alpha_n\mu\nu \alpha_1 \lambda}
+
\frac{1}{2} R_{\alpha_1\lambda \alpha_3\cdots\alpha_n\mu\nu \alpha_2 \lambda}
+ \cdots
\\&\quad%
+
\frac{1}{2} R_{\alpha_1\cdots\alpha_{n-1}\lambda\mu\nu \alpha_n \lambda}
.
\label{eq:B8}
\ees%

Eventually we will need to take traces of operators with a factor $Z^R_I$ on
the left. The formulas for the coordinate-scalar and -vector spaces are,
respectively,
%
\bes%
\tr_0( Z^R_I \cO )
&%
= - \tr_0( C_I \cO )
,
\qquad
\\
\tr_1( Z^R_I \cO^\mu{}_\nu )
&%
=  \tr_1 \left( (R_I{}^\mu{}_\lambda
-g^\mu{}_\lambda C_I ) \cO^\lambda{}_\nu \right)
.
\label{eq:8.17}
\ees%
Further details and proofs are given in App. \ref{app:a}.

\subsection{\textsf{ Integrals and traces  }
\label{sec:31c}}

We will use the shorthand notation
\be
\esp{X}_x \equiv \int d^dx\,\sqrt{g} \,X
,
\label{eq:8.32}
\ee
as well as
\be
\esp{X}_{x,g} \equiv
\int d^dx\,\sqrt{g} \,\tr_g( X )
= \esp{\tr_g( X ) }_x
,
\label{eq:8.32a}
\ee
where $\tr_g(~)$ refers to trace over gauge space. In particular $\tr_g(1)=N$,
the dimension of the gauge space is the number of dynamical real vector fields
in the theory.

In addition, for integrals over a momentum variable in Sec. \ref{sec:12},
\be\esp{X}_p \equiv
\frac{1}{\sqrt{g}} \int \frac{d^{d+2\epsilon}p}{(2\pi)^d} \, X
,
\label{eq:8.32c}
\ee
where $d+2\epsilon$ refers to dimensional regularization.
We also use combination such as $\esp{X}_{x,p}$ for $\esp{\esp{X}_p}_x$, etc.

\section{\textsf{ Results for $\Gammadiv_1$
}
\label{sec:9}}

\subsection{\textsf{ Results for $\Gamma_\gh$ and $\Gamma_{L,1}$
}
\label{sec:9a}}

The contributions $\Gamma_\gh$ and $\Gamma_{L,1}$ to the one-loop effective
action are given by Eqs. \nec{6.8} and \nec{7.7}, respectively.  The
computation of their UV divergent part is straightforward in dimension
regularization, in $d+2\epsilon$ dimensions, using the identity
\be
\Tr\,\log(-\nabla^2)\big|_{\div} =
\frac{1}{(4\pi)^{d/2}}\frac{1}{\epsilon} \int d^dx \, \sqrt{g}\, \tr(b_{d/2}(x))
,
\label{eq:9.1}
\ee
where the trace is taken in the corresponding space and $b_n$ is the $n$-th
heat-kernel coefficient of the Laplacian \cite{Vassilevich:2003xt}. For $d=2$
and $d=4$ the required coefficients are
%
\bes%
b_1
&%
= \frac{1}{6}\bR, \qquad
\\%
b_2
&%
= \frac{1}{12} Z^2_{\mu\nu}
+ \frac{1}{180} R_{\mu\nu\alpha\beta}^2
- \frac{1}{180} \cR_{\mu\nu}^2
+ \frac{1}{72} \bR^2
\,.
\ees%
As they stand, these formulas hold for an arbitrary space since $Z_{\mu\nu}$
takes care of all required curvatures (coordinate, gauge or other in more
general cases). For the space of coordinate tensors of rank $r$ in $d$
dimensions (and adjoint gauge representation) one easily finds
%
\bes%
\tr_r(Z^2_{\mu\nu})
&%
= \tr_r(F^2_{\mu\nu})
+ \tr_r((Z^R_{\mu\nu})^2)
\\&%
= d^r \tr_g(F^2_{\mu\nu})
- r d^{r-1} N R_{\mu\nu\alpha\beta}^2
,
\ees%
where as already said $\tr_g(~)$ denotes the trace over gauge space. Of course
this result is fully consistent with \eq{8.17}.

Therefore, for $d=2$,
\bes
\Gamma_\gh^\div &= -\frac{1}{4\pi\, \epsilon}
\Esp{
  \frac{1}{6}\, \bR
}_{x,g}
\,,
\\
\Gamma_{L,1}^\div &=
\frac{1}{4\pi\, \epsilon}
\Esp{
\frac{1}{6}\,\bR
}_{x,g}
\,.
\label{eq:9.4}
\ees
These two contributions cancel each other, as they should in $d=2$.

For $d=4$,
%
\bes
\Gamma_\gh^\div &= -\frac{1}{(4\pi)^2\, \epsilon}
\Bigg\langle
\frac{1}{12} F_{\mu\nu}^2
\\&\quad
+ 
\frac{1}{180} R_{\mu\nu\alpha\beta}^2
- \frac{1}{180} \cR_{\mu\nu}^2
+ \frac{1}{72} \bR^2
\Bigg\rangle_{x,g}
,
\\
\Gamma_{L,1}^\div &=
\frac{1}{(4\pi)^2\, \epsilon}
\Bigg\langle
\frac{1}{6} F_{\mu\nu}^2
\\&\quad
+
-\frac{11}{360} R_{\mu\nu\alpha\beta}^2
- \frac{1}{90} \cR_{\mu\nu}^2
+ \frac{1}{36} \bR^2
\Bigg\rangle_{x,g}
.
\ees
%

\subsection{\textsf{ Results for $\Gamma_S$}
\label{sec:3}}

\subsubsection{\textsf{Contributions to $\Gamma_S$
}}

The UV divergent part of $\Gamma_S$ in $d$ dimensions is contained in
$\sum_{n=1}^d\Gamma_{S,n}$ where $\Gamma_{S,n}$ is given in \Eq{7.8}. The
quantity $K_L$, defined in \Eq{7.6}, is homogeneous in $\nabla$ of degree $+2$
while $K_S\equiv K-K_L$ contains terms of degrees $0$ and $1$. We will expand
$\Gamma_{S,n}$ in powers of $\nabla$ keeping terms up to $O(\nabla^{-4})$
which is sufficient for $\Gamma_1^\div$ in spacetime dimensions $d \le 4$.
The trace cyclic property can be used to collect equivalent terms and also we
choose whenever possible to bring the trace to the scalar-coordinate space.

Introducing the notation
\be
\Deltag \equiv \frac{1}{\nabla^2},
\qquad
\DeltaM \equiv \frac{1}{\nabla_M^2}
,\ee
this procedure yields the following expressions
\begin{widetext}%
\begin{equation}\begin{split}
    \Gamma_{S,1} &=
    \Tr_1\Big(
    -\frac{1}{2} \Deltag\, Y_{\mu\nu}
    \Big)
    + \Tr_0\Big(
    -\frac{1}{2} \DeltaM W 
    -\frac{1}{4} \DeltaM \{ P_\mu,\nabla_\mu \}
    \Big)
    \\
    \Gamma_{S,2} &=
    \Tr_1\Big(
    -\frac{1}{4} \Deltag\, Y_{\mu\alpha} \Deltag\, Y_{\alpha\nu}
    \Big)
    +\Tr_0\Big(
     \frac{1}{2} \DeltaM\Phi_\mu \Deltag\, \Phi_\mu 
    +\frac{1}{2} \DeltaM \nabla_\mu \cM_{\mu\nu} \Deltag\, \cM_{\nu\alpha}
    \nabla_\alpha 
    \\ &
    -\frac{1}{2} \DeltaM \Phi_\mu \Deltag\, \cM_{\mu\nu} \nabla_\nu 
    -\frac{1}{2} \DeltaM \nabla_\mu \cM_{\mu\nu} \Deltag\, \Phi_\nu
    \\ &
    -\frac{1}{4} \DeltaM W\DeltaM W
    -\frac{1}{16} \DeltaM \{ P_\mu,\nabla_\mu \} \DeltaM \{
    P_\nu,\nabla_\nu \} 
    -\frac{1}{4} \DeltaM \{ P_\mu,\nabla_\mu \} \DeltaM W 
    \Big)
    \\
    \Gamma_{S,3} &=
    \Tr_0\Big(
    \frac{1}{2} \DeltaM \nabla_\mu \cM_{\mu\nu} \Deltag\, Y_{\nu\alpha}
       \Deltag\, \cM_{\alpha\beta} \nabla_\beta 
       + \frac{1}{2}  \DeltaM W \DeltaM
       \nabla_\mu \cM_{\mu\nu} \Deltag\, \cM_{\nu\alpha} \nabla_\alpha 
      + \frac{1}{4}  \DeltaM \{ P_\mu,\nabla_\mu \}
        \DeltaM \nabla_\nu \cM_{\nu\alpha}  \Deltag\, \cM_{\alpha\beta} \nabla_\beta 
       \\ &
       -\frac{1}{4}  \DeltaM \{ P_\mu,\nabla_\mu \} \DeltaM
       \nabla_\nu  \cM_{\nu\alpha} \Deltag\, \Phi_\alpha 
      -\frac{1}{4}  \DeltaM \{ P_\mu,\nabla_\mu \}
        \DeltaM \Phi_\nu \Deltag\, \cM_{\nu\alpha} \nabla_\alpha 
        \\ &
      -\frac{1}{8}  \DeltaM W \DeltaM \{ P_\mu,\nabla_\mu \}
       \DeltaM \{ P_\nu,\nabla_\nu \} 
        -\frac{1}{48}  \DeltaM \{ P_\mu,\nabla_\mu \}
        \DeltaM \{ P_\nu,\nabla_\nu \} \DeltaM \{ P_\alpha,\nabla_\alpha \} 
        + O(\nabla^{-5})
        \Big)
        \\
    \Gamma_{S,4} &=
    \Tr_0\Big(
    -\frac{1}{4} 
    \DeltaM \nabla_\mu \cM_{\mu\nu}  \Deltag\, \cM_{\nu\alpha} \nabla_\alpha
    \DeltaM \nabla_\beta \cM_{\beta\rho}  \Deltag\, \cM_{\rho\sigma}
    \nabla_\sigma 
    +\frac{1}{8}  \DeltaM \{ P_\mu,\nabla_\mu \}
        \DeltaM \{ P_\nu,\nabla_\nu \}
    \DeltaM \nabla_\alpha \cM_{\alpha\beta}  \Deltag\, \cM_{\beta\rho} \nabla_\rho
    \\ &
     -\frac{1}{128}  \DeltaM \{ P_\mu,\nabla_\mu \}
     \DeltaM \{ P_\nu,\nabla_\nu \} \DeltaM \{ P_\alpha,\nabla_\alpha \}
     \DeltaM \{ P_\beta,\nabla_\beta \}   
     + O(\nabla^{-5})
     \Big)
     \label{eq:8.1}
\end{split}\end{equation}
\end{widetext}%

The functional traces are of the form
\be
\Tr_r(A) = \int d^dx \sqrt{g}\, 
\tr_r( \esp{x|A|x} ),
\qquad
r=0,1
\label{eq:8.10b}
\ee
where $d$ is the spacetime dimension and $\tr_0$ or $\tr_1$ refer to the trace
over coordinate labels, scalar or vector, respectively, and also include trace
over gauge labels. In detail, the diagonal (in $x$-space) matrix element
$\esp{x|A|x}$ is of the form $\esp{x,I',a | A| x,I,b}$ where $a,b$ are gauge
labels in the adjoint representation and $I,I'$ are coordinate labels. For
$\Tr_0(~)$ these coordinate labels are absent while for $\Tr_1(~)$ they are of
the type $I=\mu$, $I'=\nu$.

At this point one could already attempt the computation the functional traces
to obtain $\Gammadiv_S$. Nevertheless it is convenient to first simplify the
expressions by bringing the operators to a canonical form. This is in the same
spirit as the universal functional traces of \cite{Barvinsky:1985an}. The
goal is to put together terms involving powers of $\nabla$ (to wit
$\nabla_\mu$, $\Delta$ and $\Delta_M$) on one side and the purely
multiplicative terms on another. That is, bring the various operators in
\nec{8.1} to the form
\be
A = \sum_n \cO_n \, A_n
\,,
\label{eq:8.4}
\ee
where $\cO_n$ form a basis of pseudo-differential operators and $A_n$ are
purely multiplicative operators. We have chosen to put the latter on the
right-hand side. Thus, for the diagonal matrix elements
\be
\esp{x,I',a | A| x,I,b}
= \sum_{n,c} \esp{x,I',a | \cO_n | x,I,c} \, (A_n(x))_{cb}
,
\ee
or just $\esp{x| A| x} = \sum_{n} \esp{x | \cO_n | x} \, A_n(x)$.  The
coefficients $A_n$ do not modify the UV degree of divergence of the term,
which is controlled by $\cO_n$, so the hardest work is computing $\esp{x |
  \cO_n | x}$.

To obtain the expansion in \Eq{8.4} we apply the following commutation
identities,
\begin{widetext}%
\bes
    [\nabla_\mu,X_I] &= X_{\mu I}
, \\
  [\Deltag,X_I] &=
  \Deltag \big( X_{\mu\mu I}
  -  2\nabla_\mu X_{\mu I} \big) \Deltag
, \\
  [\Deltag,\nabla_\mu] &=
  \Deltag \big(
  2\nabla_\nu Z_{\mu\nu}
  - \nabla_\nu \cR_{\mu\nu}
  + Z_{\nu\nu\mu}
  + \frac{1}{4} \bR_\mu \big) \Deltag
, \\
  [\DeltaM,\nabla_\mu] &=
  \DeltaM \big(
    \nabla_\alpha \nabla_\beta M_{\mu\alpha\beta}
  - \nabla_\alpha M_{\beta \mu \beta \alpha}
  + \nabla_\alpha M_{\alpha\beta} Z_{\mu\beta}
  + Z_{\mu\alpha} \nabla_\beta M_{\alpha\beta}
  - Z_{\mu \alpha} M_{\beta\beta\alpha}
  \big) \DeltaM
, \\
  [\Deltag , \ZR_I] &=
  \Deltag \big(
   \nabla_\mu R_{\nu I \mu\nu}
  - 2\nabla_\mu \ZR_{\mu I}
  + \ZR_{\mu\mu I}
  \big) \Deltag
, \\
  [\DeltaM , \ZR_I] &=
  \DeltaM \left(
    \nabla_\mu M_{\mu\nu} \big(
  R_{\alpha I \nu \alpha} - 2 \ZR_{\nu I} \big)
  + M_{\mu\nu} \big( \ZR_{\mu\nu I} - \frac{1}{2} R_{\mu\alpha I \nu\alpha}
  - \frac{1}{2} R_{\alpha\mu I \nu\alpha} \big)
  +  M_{\mu\mu\nu} \big( \ZR_{\nu I} -\frac{1}{2} R_{\alpha I \nu\alpha} \big)
  \right) \DeltaM
  .
\label{eq:8.3}
\ees
\end{widetext}%

In these formulas $I=\mu_1\cdots\mu_n$\/ represents any (possibly empty) string of coordinate
indices (e.g. $\alpha\beta$) and $\mu I$ the new string adding $\mu$ to the
left (e.g. $\mu\alpha\beta$). $X_I$ represents a purely multiplicative
operator, that is, any coordinate tensor (in particular may be a coordinate
scalar) and a matrix in gauge space, not involving $\nabla_\mu$ nor
$Z_I^R$.

The usefulness of the commutation relations \nec{8.3} is that the true degree
of UV divergence of the operator on the left-hand side is actually smaller
than the nominal one. So for instance, $[\Deltag,X_I]$ would be nominally of
$O(\nabla^{-2})$ while this commutator is actually of order
$O(\nabla^{-3})$. Note that $Z_I$ and $Z^R_I$ count as degree $O(\nabla^0)$ as
they do not add to the UV degree of divergence of a term, as follows from
the property in \Eq{8.6a}.

A very conspicuous and relevant absence in the list of commutators is the
combination $[ \DeltaM, X_I]$. This is not in the list because that commutator
is still of $O(\nabla^{-2})$ unless $[X_I,M_{\mu\nu}]=0$. This absence
prevents from putting together all terms involving $\nabla$ and is a consequence of the
non-Abelian character of the theory.

For $\Gamma_S$, the operators required in the basis are
\begin{widetext}%
\be
\begin{aligned}
  (\cO_1)_\mu &= \DeltaM \nabla_\mu
&
  \cO_2 &= \DeltaM
\\
  (\cO_3)_{\mu\nu} &= \DeltaM \Deltag \nabla_\mu \nabla_\nu
&
(\cO_4[X])_{\mu\nu} &= \DeltaM X \DeltaM \nabla_\mu \nabla_\nu
\\
(\cO_5)_\mu &= \DeltaM \Deltag \nabla_\mu
&
(\cO_6)_{\mu\nu\alpha} &= \DeltaM \Deltag^2
\nabla_\mu \nabla_\nu \nabla_\alpha
\\
(\cO_7[X])_\mu &= \DeltaM X \DeltaM \nabla_\mu
&
(\cO_8[X])_{\mu\nu\alpha} &= \DeltaM X \DeltaM
\Deltag \nabla_\mu \nabla_\nu \nabla_\alpha
\\
(\cO_9[X,X'])_{\mu\nu\alpha} &= \DeltaM X \DeltaM X'
\DeltaM \nabla_\mu \nabla_\nu \nabla_\alpha
&
\cO_{10} &= \DeltaM \Deltag
\\
(\cO_{11})_{\mu\nu} &= \DeltaM \Deltag^2
\nabla_\mu \nabla_\nu
&
(\cO_{12})_{\mu\nu\alpha\beta} &= \DeltaM \Deltag^3 
\nabla_\mu \nabla_\nu \nabla_\alpha \nabla_\beta
\\
\cO_{13}[X] &= \DeltaM X \DeltaM
&
(\cO_{14}[X])_{\mu\nu} &= \DeltaM X \DeltaM
\Deltag \nabla_\mu \nabla_\nu
\\
(\cO_{15}[X])_{\mu\nu\alpha\beta} &= \DeltaM X \DeltaM \Deltag^2
\nabla_\mu \nabla_\nu \nabla_\alpha \nabla_\beta
&
(\cO_{16}[X,X'])_{\mu\nu} &= \DeltaM X \DeltaM X'
\DeltaM \nabla_\mu \nabla_\nu
\\
(\cO_{17}[X,X'])_{\mu\nu\alpha\beta} &= \DeltaM X
\DeltaM X' \DeltaM \Deltag
\nabla_\mu \nabla_\nu \nabla_\alpha \nabla_\beta
&
(\cO_{18}[X,X',X''])_{\mu\nu\alpha\beta} &= \DeltaM X
\DeltaM X'  \DeltaM X'' \DeltaM
\nabla_\mu \nabla_\nu \nabla_\alpha \nabla_\beta
\\
\cO_{19} &= \Deltag
&
\cO_{20} &= \Deltag^2
\end{aligned}
\label{eq:8.6}
\ee
\end{widetext}%
Here $X,X',X''$ are arbitrary purely multiplicative operators, possibly with
coordinate indices. The presence of operators in the basis with such
insertions of purely multiplicative operators is a direct consequence of
$[\DeltaM,X]$ being of $O(\nabla^{-2})$ in the non-Abelian case.  Nevertheless,
the computation of the diagonal matrix elements of the operators $\cO_n$ can
be done for generic $X,X',X''$.

The operators $\cO_1$ through $\cO_{18}$ are required for the terms $\Tr_0(~)$
of $\Gamma_S$ in $d=4$ dimensions, $\cO_{19}$ and $\cO_{20}$ appear in the
terms $\Tr_1(~)$. The explicit expansion of the operators in $\Gamma_S$,
\Eq{8.1}, in the basis \nec{8.6} is presented in App. \ref{app:c}.

The traces indicated in \Eq{8.1} can then be obtained from
\be
\Tr_r(A) = \left\langle \sum_n 
  \tr_r( \esp{x|\cO_n|x} \, A_n(x))
  \right\rangle_x
  ,
  \qquad
r=0,1
\,.
\label{eq:8.10a}
\ee

The diagonal matrix elements of the operators $\cO_n$ can be regarded as a
generalization of the well-known {\em universal functional traces} introduced
in \cite{Barvinsky:1985an} and computed there using a Schwinger-DeWitt
technique. The same technique was adapted in \cite{Ruf:2018vzq} for the
Abelian case, where $M^{\mu\nu}$ acts effectively as a second metric. Our
problem involves considerably more complicated operators since now such
effective metric is gauge non-Abelian. We will use the method of covariant
symbols, already applied in \cite{Garcia-Recio:2019iia} in the Abelian
setting. Of course, it would be interesting to adapt the Schwinger-DeWitt
approach to the present non-Abelian case.

\subsubsection{\textsf{$\Gamma_S^\div$ in $d=2$}}

\begin{table}[t]
\centering
\begin{tabular}{r r r r r} 
  \hline
 Operator & degree & $-2$ & $-4$ & $-6$ 
 \\ [0.5ex] 
 \hline
 $\cO_1$   & $-1$ & 2 & 265 & 
 \\
 $\cO_2$   & $-2$ & 1 & 11  & 3303
 \\
 $\cO_3$   & $-2$ & 1 & 44 &   
 \\
 $\cO_4$   & $-2$ & 1 & 123 & 
 \\
 $\cO_{5,6}$   & $-3$ &   &  2 &  
 \\
 $\cO_{7,8}$   & $-3$ &   &  6 &
 \\
 $\cO_9$   & $-3$ &   &  12 &   
 \\
 $\cO_{10\,\text{-}18}$ & $-4$ &   &  1 &
 \\
 $\cO_{19}$ & $-2$ & 1 &  1 &
 \\
 $\cO_{20}$ & $-4$ &   &  1 &  
 \\ 
 \hline
\end{tabular}
\caption{For the various operators in \Eq{8.6}, the second column displays the
  UV divergence degree, while the columns labeled with $-2,-4,-6$, show the
  number of terms of each degree in the expansion of the diagonal matrix
  element of the operator.}
\label{tab:1}
\end{table}

Details of the calculation of the divergent part of the diagonal matrix
elements will be given in Sec. \ref{sec:12}. However, from the data displayed
in Table \ref{tab:1}, we already anticipate that the number of terms
contributing to $\Gamma_S^\div$ is very large for $d=4$ (namely, $265$ terms
from $\cO_1$, $123$ from $\cO_4$, etc). In view of this we will only display
explicit results for $d=2$.

In $d=2$, \Eq{8.1} reduces to
%
\begin{equation}\begin{split}
    \Gamma_{S,1} &=
    \Tr_1\Big(
    -\frac{1}{2} \Deltag\, Y_{\mu\nu}
    \Big)
    + \Tr_0\Big(
    -\frac{1}{2} \DeltaM W 
    -\frac{1}{4} \DeltaM \{ P_\mu,\nabla_\mu \}
    \Big)
    \\
    \Gamma_{S,2} &=
    \Tr_0\Big(
    \frac{1}{2} \DeltaM \nabla_\mu \cM_{\mu\nu} \Deltag\, \cM_{\nu\alpha}
    \nabla_\alpha 
    \\ &\quad
    -\frac{1}{16} \DeltaM \{ P_\mu,\nabla_\mu \} \DeltaM \{ P_\nu,\nabla_\nu
    \}
    + O(\nabla^{-3})
    \Big)
\end{split}\end{equation}
This can be rewritten using the basis of operators $\cO_n$ in \Eq{8.6} (see
App. \ref{app:c})
\begin{equation}\begin{split}
    \Gamma_{S,1} &=
    \Tr_1\Big(
    -\frac{1}{2} \cO_{19}\, Y_{\mu\nu}
    \Big)
    + \Tr_0\Big(
    -\frac{1}{2} \cO_2 W 
    -\frac{1}{2} (\cO_1)_\mu P_\mu
    \\ &\quad
    + \frac{1}{4} \cO_2 \,P_{\mu\mu}
    + O(\nabla^{-3})
    \Big)
    \\
    \Gamma_{S,2} &=
    \Tr_0\Big(
    \frac{1}{2} (\cO_3)_{\mu\nu} \cM_{\mu\alpha} \cM_{\alpha\nu}
    -\frac{1}{4} (\cO_4[P_\mu])_{\mu\nu} \, P_\nu
    + O(\nabla^{-3})
    \Big)
    \label{eq:10.11a}
\end{split}\end{equation}
%

Details of the calculation are given in Sec. \ref{sec:12.2}. The result
is\mfootnote{Note that
  $ Y_{\mu\mu} = M_{\mu\mu} + \bR, \quad P_{\mu\mu} =
  \frac{1}{2}[F_{\mu\nu},Q_{\mu\nu}] $.}
%
\bes
\Gamma_{S,1}^\div &=
\frac{1}{4\pi \epsilon}\Big\langle 
- \frac{1}{2} Y_{\mu\mu}
- \frac{1}{2} \hat{N}_M W
+ \frac{1}{4} \hat{N}_M P_{\mu\mu}
\\&\quad
- \frac{1}{2} \hat{N}_M M_{\mu\nu} \hat{N}_M M_{\mu\alpha\beta} \hat{N}_M P_\lambda
\hk_\nu \hk_\alpha \hk_\beta \hk_\lambda
\\&\quad
+ \frac{1}{2} \hat{N}_M M_{\mu\alpha\beta} \hat{N}_M M_{\mu\nu} \hat{N}_M P_\lambda
\hk_\nu \hk_\alpha \hk_\beta \hk_\lambda
\Big\rangle_{x,\, g, \,\ang,}
,
\\
\Gamma_{S,2}^\div &=
\frac{1}{4\pi \epsilon}\Big\langle
\frac{1}{2} \hat{N}_M \cM_{\mu\alpha}\cM_{\alpha\nu} \, \hk_\mu \hk_\nu
- \frac{1}{4} \hat{N}_M P_\mu \hat{N}_M P_\nu \hk_\mu \hk_\nu
\Big\rangle_{x,\, g, \,\ang}
\,.
\label{eq:8.36b}
\ees
Here $\hk_\mu$ is a normalized momentum variable,
$g^{\mu\nu} \hk_\mu\hk_\nu=1$, whereas
\be
\hat{N}_M
\equiv ( \hk_\mu \hk_\nu M^{\mu\nu} )^ {-1}
\,.
\ee
The symbol $\esp{~}_{\ang}$ denotes angular average over $\hk_\mu$,
\be
\esp{ X }_{\ang}
\equiv
\frac{\Gamma(d/2)}{2\pi^{d/2}}
\int d^{d-1}\Omega_{\hk} \, X
\,.
\ee
This average is to be applied together with $\esp{~}_{x,g}$, already introduced
in Sec. \ref{sec:31c}.

\Eq{8.36b} is our final result for $\Gamma^\div_S$ in $d=2$. The expression for
$d=4$ is qualitatively similar, but considerably longer. Further perturbative
results are given in App. \ref{app:d}.

While the angular averages in \Eq{8.36b} cannot be evaluated in closed form in
general, they are perfectly convergent and well-defined. In any case, these
integrals introduce a wild local but non polynomial dependence on the field
$M^{\mu\nu}_{ab}(x)$ in the divergent part of the effective action, implying
that the UV divergences cannot be removed by polynomial counterterms,
rendering the theory not renormalizable in a standard sense. This was true
already when $N=1$, the Abelian case studied in
\cite{Toms:2015fja,Buchbinder:2017zaa,Ruf:2018vzq,Garcia-Recio:2019iia}.

\subsection{\textsf{ Results for $\Gamma_{L,0}$}
  \label{sec:10}}

\subsubsection{\textsf{ Preliminaries }
  \label{sec:10.1}}

The remaining contribution to the effective action is $\Gamma_{L,0}$,
\be
\Gamma_{L,0} = \frac{1}{2}\Tr_0\log(-\nabla_M^2)
= \frac{1}{2}\Tr_0\log(-\nabla_\alpha M^{\alpha\beta} \nabla_\beta)
.
\ee
This is the effective action of a scalar field with action
\be
S_{L,0} = \int d^dx\,\sqrt{g}\, \frac{1}{2} \phi_\mu^a \,M^{\mu\nu}_{ab} \phi^b_\nu
.
\label{eq:10.2}
\ee
In the special case of $N=1$, that is, when there is no gauge sector (the case
considered in
\cite{Toms:2015fja,Buchbinder:2017zaa,Ruf:2018vzq,Garcia-Recio:2019iia}) one
can combine $g_{\mu\nu}$ and $M^{\alpha\beta}$ into a new metric
$\tilde{g}_{\mu\nu}$, namely, $\sqrt{g}\, M^{\mu\nu}=\sqrt{\tilde{g}}\,
\tilde{g}^{\mu\nu}$ \cite{deRham:2014wfa,Ruf:2018vzq}, in such a way that
\be
S_{L,0}
= \int d^dx\,\sqrt{\tilde{g}}\, \frac{1}{2} \phi_\mu
\,\tilde{g}^{\mu\nu} \phi_\nu
\qquad (N=1)
.
\ee
Hence $\Gamma_{L,0} = \frac{1}{2}\Tr_0\log(-\tilde{\nabla}^2)$ and the
heat-kernel result \nec{9.1} immediately applies. Note that this method works
in any spacetime dimension except $d=2$, since in that case
$\det(\sqrt{\tilde{g}}\tilde{g}_{\mu\nu}) \equiv 1$.

Unfortunately no such simplification takes place for $N>1$.  In fact the
situation is even worse, namely, $\Gammadiv_{L,0}$ does not admit a standard
form, like that displayed in \Eq{8.36a}. By definition, we say that the terms
in an expansion of a diagonal matrix element $\esp{x|\cO|x}$ adopt a {\em
  standard form} when all the pieces are covariant with no ``free''
$\nabla_\mu$ operators (all $\nabla_\mu$ are in the form $[\nabla_\mu,~]$) and
only remains to carry out a momentum integration. Also labeled operators
\cite{Salcedo:2004yh} are not permitted in a standard form. The momentum
integration can be traded by another parameter, e.g. a proper time as in a
Schwinger parametrization.

While a standard form can always be achieved for matrix elements of the type
$\esp{x|f(\nabla,X)|x}$ ($X$ being purely multiplicative operators) when the
dependence on $\nabla_\mu$ is of rational type, this is not guaranteed for
more general functions $f$. However, it is often the case that a standard form
exist for a pseudo-differential operator of the type $\log(\hat{f})$ with
$\hat{f}$ a differential operator. A well-known instance of standard form for
a $\Tr\log$ is that found by Chan (for flat spacetime to fourth order in a
derivative expansion) \cite{Chan:1986jq}. To second order
\bes
\Tr\log(-\nabla_\mu^2 + X) &=
\\ & \hspace{-36pt}
\Esp{ \int\frac{d^dk}{(2\pi)^d} \left(
  -\log(N_C) + \frac{k^2}{d} [\nabla_\mu,N_C]^2 + O(\nabla^4)
\right)
}_{x,g}
,
\qquad
\\
N_C &\equiv (k^2 + X)^{-1}
.
\ees
Extensions of this formula exist to sixth order \cite{Caro:1993fs}, also for
curved spacetime \cite{Salcedo:2007bt} and for finite temperature
\cite{Moral-Gamez:2011wcb}. Yet the techniques applied in those cases cannot
be translated to evaluating $\Gamma_{L,0}$ because in the present case what
plays the role a metric, $M^{\mu\nu}$, is actually a non-trivial matrix in
gauge space and appears already in the leading term
$\log(k_\mu k_\nu M^{\mu\nu})$. The obstruction to a standard form does not
depend on the method used to compute the effective action, it is intrinsic to
$\Gamma_{L,0}$ in the non-Abelian case.

\subsubsection{\textsf{Special case of separable mass term}
  \label{sec:10.A1}}

Here we make an apart to mention the special case of a mass term separable in
coordinate  and gauge spaces, that is,
\be
M^{\mu\nu}_{ab}(x) = T_{ab}(x) G^{\mu\nu}(x)
\label{eq:10.6}
.
\ee
The fluctuation operator can then be written as
%
\bes%
\nabla_\alpha M^{\alpha\beta} \nabla_\beta
&%
=
\nabla_\alpha T G^{\alpha\beta} \nabla_\beta =
T \nabla_\alpha G^{\alpha\beta} \nabla_\beta
+
T_\alpha  G^{\alpha\beta} \nabla_\beta
\\&
=
T(\nabla_G^2 + L^\alpha \, \nabla_\alpha)
\ees
where we have defined
\be
\nabla_G^2 \equiv \nabla_\alpha G^{\alpha\beta} \nabla_\beta,
\qquad
L^\mu \equiv T^{-1} T_\nu \, G^{\nu\mu}
.
\ee
Now, the factor $T(x)$ can be dropped from $\Tr(\log(-\nabla_M^2))$, being
ultralocal (carries no derivative operators) hence
%
\bes%
\Gamma_{L,0}
&%
=
\frac{1}{2} \Tr_0 \log(- \nabla_G^2 - L^\alpha \nabla_\alpha )
\\&%
=
\frac{1}{2} \Tr_0 \log( - \nabla_G^2 )
+
\frac{1}{2} \Tr_0 \log( 1 + \Delta_G \, L^\alpha \nabla_\alpha )
\label{eq:10.9a}
\ees%
with $\Delta_G = 1/\nabla_G^2$. The first term is just like $\Gamma_{L,0}$ of
the Abelian case $N=1$ (with $G^{\mu\nu}$ instead of $M^{\mu\nu}$). The second
term can be expanded in powers of $\Delta_G \, L^\alpha \nabla_\alpha$,
similarly as done for $\Gamma_S$ in \nec{7.8a}. So this contribution does
admit a standard form.

It is noteworthy that there is an ambiguity in the separation \nec{10.6} of
Weyl-transformation type, namely,
\be
G^{\mu\nu}(x) \to \lambda(x) G^{\mu\nu}(x), \qquad
T_{ab}(x) \to \lambda^{-1}(x) T_{ab}(x)
,
\ee
where $\lambda(x)$ is local but both scalar coordinate and gauge singlet. Such
ambiguity can be used to fix some gauge condition on $G^{\mu\nu}(x)$ or
$T_{ab}(x)$, or as a check of the calculation, since the sum of the two terms
in \nec{10.9a} should be $\lambda$-independent.

 We do not pursue the subject of the separable mass case any further in this
 work.

\subsubsection{\textsf{Method of contour integration}
\label{sec:10.B}}

Coming back to the case of a general mass term $M^{\mu\nu}_{ab}(x)$, to deal
with the logarithm we follow here the approach of introducing a parametric
integral. This is based on the observation that a standard form would easily
follow for a rational function instead of the $\log$.

The contour integration method is based on the identity
\be
\Gamma_{L,0} =
\int_\gamma \frac{dz}{2\pi i} \, \log(z)
\,\frac{1}{2}\Tr_0\left(\frac{1}{z + \nabla_M^2}\right)
\label{eq:10.11}
\ee
where the path $\gamma$ is meant to enclose counterclockwise the spectrum of
$-\nabla_M^2$. More precisely, the path $\gamma$ on the $z$ complex plane
starts at $-\infty$ towards the origin just above the negative real semiaxis,
encircles $z=0$ clockwise and goes back to $-\infty$ just below the negative
real semiaxis.\mfootnote{The $\zeta$-function result would follows from
  understanding $\log(z)$ as $-\frac{d z^{-s}}{ds}$ at $s=0$ in the sense of
  analytical continuation in the $s$ variable, taking the derivative with
  respect to $s$ only after the $\Tr(~)$ has been computed
  \cite{Seeley:1967ea}.} Hence, introducing a convenient notation,
\be
\esp{X}_z \equiv
\int_\gamma \frac{dz}{2\pi i} \, \log(z)
\, X
,
\qquad
\cO_2' \equiv \frac{1}{z + \nabla_M^2}
,
\label{eq:10.8}
\ee
we can express \Eq{10.11} as
\be
\Gamma_{L,0} =\Esp{
  \frac{1}{2}\Tr_0 (\cO_2')
}_z
.
\label{eq:10.8a}
\ee

The calculation of the diagonal matrix element $\esp{x|\cO_2'|x}$ is formally
identical to that of $\esp{x|\cO_2|x}$, unfortunately the presence of the
$z$-integration implies that relevant quantity
$\esp{\cO_2'}_z=\log(-\nabla_M^2)$ is actually of degree $O(\nabla^0)$
instead of $O(\nabla^{-2})$. Hence, the amount of work required to evaluate
$\esp{x|\esp{\cO_2'}_z|x}$, is similar to that of $\esp{x|\cO_2|x}$ for $d+2$
dimensions to achieve the same degree of UV divergence (see Sec. \ref{sec:12.3}).

The convergence problem would not improve if instead of a $z$-integration the
sought-for rational dependence on $\nabla_\mu$ were obtained through the
variation of $\Gamma_{L,0}$ under a deformation of $M^{\mu\nu}$,
\be
\delta \Gamma_{L,0} = \frac{1}{2} \Tr_0
\left( \DeltaM \nabla_\mu  \delta  M^{\mu\nu} \nabla_\nu
\right)
,
\ee
as this requires to evaluate $\esp{x| \Delta_M \nabla_\mu \nabla_\nu| x}$ which
is still of $O(\nabla^0)$.

\subsubsection{\textsf{$\Gamma_{L,0}$ in $d=2$}
\label{sec:10.C}}

Once again the number of terms is prohibitively large in $d=4$ (it requires
$\cO_2$ to $O(p^{-6})$ in Table \ref{tab:1}) and we present results for
$d=2$. The calculation gives
%
\bes
&\Gamma_{L,0}^\div
=
\frac{1}{4\pi \epsilon}\Big\langle
\\&\quad
-  \hat{N}_M \,M_{\mu  \nu }\, \hat{N}_M \,M_{\mu  \alpha  \beta
   }\, \hat{N}_M \,M_{\rho  \sigma }\, \hat{N}_M \,M_{\rho  \lambda  \eta }\, \hat{N}_M \,\hk_{\nu
}\,\hk_{\alpha }\,\hk_{\beta }\,\hk_{\sigma }\,\hk_{\lambda }\,\hk_{\eta }
\\&\quad
-  \hat{N}_M \,M_{\mu  \nu
    \alpha }\, \hat{N}_M \,M_{\mu  \beta }\, \hat{N}_M \,M_{\rho  \sigma  \lambda
   }\, \hat{N}_M \,M_{\rho  \eta }\, \hat{N}_M \,\hk_{\nu }\,\hk_{\alpha }\,\hk_{\beta }\,\hk_{\sigma
}\,\hk_{\lambda }\,\hk_{\eta }
\\&\quad
+\frac{1}{2} \hat{N}_M \,M_{\mu  \nu }\, \hat{N}_M \,M_{\mu  \alpha  \beta
   }\, \hat{N}_M \,M_{\alpha  \rho  \sigma }\, \hat{N}_M \,\hk_{\nu }\,\hk_{\beta }\,\hk_{\rho }\,\hk_{\sigma
}
\\&\quad
- \frac{1}{2} \hat{N}_M \,M_{\mu  \nu }\, \hat{N}_M \,M_{\alpha  \alpha  \beta }\, \hat{N}_M \,M_{\mu  \rho
   \sigma }\, \hat{N}_M \,\hk_{\nu }\,\hk_{\beta }\,\hk_{\rho }\,\hk_{\sigma }
\\&\quad
+ \frac{1}{2} \hat{N}_M \,M_{\mu  \nu
    \alpha }\, \hat{N}_M \,M_{\beta  \mu  \rho }\, \hat{N}_M \,M_{\beta  \sigma }\, \hat{N}_M \,\hk_{\nu
}\,\hk_{\alpha }\,\hk_{\rho }\,\hk_{\sigma }
\\&\quad
- \frac{1}{2} \hat{N}_M \,M_{\mu  \nu  \alpha }\, \hat{N}_M \,M_{\beta
    \beta  \rho }\, \hat{N}_M \,M_{\mu  \sigma }\, \hat{N}_M \,\hk_{\nu }\,\hk_{\alpha }\,\hk_{\rho
}\,\hk_{\sigma }
\\&\quad
- \frac{1}{2} \hat{N}_M \,M_{\mu  \nu }\, \hat{N}_M \,M_{\alpha  \beta
   }\, \hat{N}_M \,M_{\mu  \rho }\, \hat{N}_M \,\bR \,\hk_{\nu }\,\hk_{\alpha }\,\hk_{\beta }\,\hk_{\rho
}
\\&\quad
+ \frac{1}{2} \hat{N}_M \,M_{\mu  \mu  \nu }\, \hat{N}_M \,M_{\alpha  \alpha  \beta
}\, \hat{N}_M \,\hk_{\nu }\,\hk_{\beta }
\\&\quad
- \frac{1}{2} \hat{N}_M \,M_{\mu  \nu }\,F_{\mu  \alpha }\, \hat{N}_M \,M_{\alpha  \beta }\, \hat{N}_M \,\hk_{\nu
}\,\hk_{\beta }
\\&\quad
- \frac{1}{2} \hat{N}_M \,M_{\mu  \nu }\, \hat{N}_M \,F_{\mu  \alpha }\,M_{\alpha  \beta
}\, \hat{N}_M \,\hk_{\nu }\,\hk_{\beta }
\\&\quad
+\frac{1}{3}  \hat{N}_M \,M_{\mu  \nu }\, \hat{N}_M \,M_{\mu  \alpha }\, \hat{N}_M \,\bR\,\hk_{\nu
}\,\hk_{\alpha }
\\&\quad
+\frac{1}{24}  \hat{N}_M \,M_{\mu  \mu }\, \hat{N}_M \,\bR
\Big\rangle_{x,g,\ang,z}
\label{eq:10.18b}
\ees
%
In this formula $\hat{N}_M \equiv (-z + \hk_\mu \hk_\nu M^{\mu\nu}
)^{-1}$. Besides integration over $x$, trace over gauge space and angular
average, an integral over $z$, as defined in \Eq{10.8}, is applied.

Integration by parts with respect to $x$ has been used to have at most one
covariant derivative on $M_{\mu\nu}$. Also the following two-dimensional
identities have been used:
\bes
R_{\mu\nu\alpha\beta} &= \frac{1}{2} ( g_{\mu\alpha} \, g_{\nu\beta} -
g_{\mu\beta} \, g_{\nu\alpha} ) \bR,
\qquad
\\
\cR_{\mu\nu} &= \frac{1}{2} g_{\mu\nu}  \bR
\,.
\label{eq:10.19}
\ees
In addition integration by parts identities in momentum space have been
applied to bring the expression to a manifestly Hermitian form.

Perturbative expansions are presented in App. \ref{app:d}.

\subsubsection{\textsf{Verification of metric-related symmetries in
    $\Gamma_{L,0}$}
  \label{sec:10.D}}

The Weyl-like transformation noted at the end of Sec. \ref{sec:2.1}, 
$g_{\mu\nu}(x) \to \xi(x) g_{\mu\nu}(x)$,
$\cM^{\mu\nu}_{ab}(x)\to \xi^{-2}(x)\cM^{\mu\nu}_{ab}(x)$, is a symmetry of the
full action only in $d=4$. However, in any spacetime dimension, $\Gamma_{L,0}$
has a large symmetry as this term only depends on the pair
$(g_{\mu\nu},M^{\mu\nu})$ through the combination $\sqrt{g}M^{\mu\nu}_{ab}$. This
follows from \nec{10.2} since $\phi^a$ are coordinate scalars and so
$\phi_\mu^a$ does not depend on the metric tensor.

It is convenient to distinguish two types of symmetry transformations leaving
invariant $\Gamma_{L,0}$, which will be called {\em transverse} and {\em
  longitudinal}, respectively:
\begin{itemize}
\item[$i)$] A transverse transformation corresponds to leaving
$M^{\mu\nu}$ intact, while $g_{\mu\nu}$ changes arbitrarily but keeping
$\sqrt{g}$ invariant, thus preserving the volume element.\\
\item[$ii)$] A longitudinal transformations correspond to the simultaneous
  change
\bes
g_{\mu\nu}(x) &\to \xi(x) g_{\mu\nu}(x),
\\
M^{\mu\nu}_{ab}(x) &\to \xi^{-d/2}(x)M^{\mu\nu}_{ab}(x),
\ees
in $d$ spacetime dimensions.
\end{itemize}

We have checked that both symmetries are fulfilled by the expression in
\Eq{10.18b}. It is sufficient to consider the infinitesimal
case.\mfootnote{Here we need to refer to results to be established in
  Sec. \ref{sec:12.3}.} For {\em transverse} transformations the first order
variations are
\bes
\delta g_{\mu\nu} &= \omega_{\mu\nu}
\quad
\text{with}\quad \omega^\lambda{}_\lambda=0,
\\
\delta M^{\mu\nu} &= \delta p_\mu = \delta F_{\mu\nu} = \delta N_M=0
,
\\
\delta M_\alpha{}^{\mu\nu} &=
\frac{1}{2}(\omega_\alpha{}^\mu{}_\lambda
+\omega_\lambda{}^\mu{}_\alpha -\omega^\mu{}_{\alpha\lambda} ) M^{\lambda\nu}
\\&\quad
+
\frac{1}{2}(\omega_\alpha{}^\nu{}_\lambda
+\omega_\lambda{}^\nu{}_\alpha -\omega^\nu{}_{\alpha\lambda} ) M^{\mu\lambda}
,
\\
\delta (g_{\mu\nu} \bR) &=
\omega^\lambda{}_{\mu \lambda \nu }
+ \omega^\lambda{}_{\nu \lambda \mu }
- \omega^\lambda{}_{\lambda \mu \nu} \quad (d=2)
.
\ees
It should be noted that the metric does not appear explicitly anywhere in
\Eq{10.18}, and the metric is only contained in $R_{\mu\nu}{}^\alpha{}_\beta$,
$\cR_{\mu\nu}$ and the covariant derivatives. This is because the same is true
for the covariant symbols of $\nabla_\mu$ and the multiplicative operators,
and no explicit metric appears in the original expression of $\cO_2'$. After
using the two-dimensional identities \nec{10.19} the metric does enter in
\Eq{10.18a} in the combination $g_{\mu\nu} \bR$. In $d=2$ the variation of
this quantity depends only on {\em derivatives} of $\omega_{\mu\nu}$. Also
note that the integral $\esp{~}_{x,p}$ does not contain a net $\sqrt{g}$. When
the above first order variations are applied to \nec{10.18a}, and after
integration by parts in $x$-space to remove terms with
$\omega_{\mu\nu\alpha\beta}$ (two derivatives), one obtains an expression
which vanishes upon integration by parts in $p$-space. We have checked this to
third order in an expansion of the type
$M^{\mu\nu}= m^2g^{\mu\nu} + H^{\mu\nu}$ in powers of $H^{\mu\nu}$. The
integration over $z$ plays no role in the cancellation of terms.

For first order {\em longitudinal} variations one has instead (in $d=2$)
\bes
\delta g_{\mu\nu} &= w g_{\mu\nu},
\quad
\delta p_\mu = \delta F_{\mu\nu} = 0
,
\quad
\delta M^{\mu\nu} = -w M^{\mu\nu}
,
\\
\delta M_\alpha{}^{\mu\nu} &=
-w M_\alpha{}^{\mu\nu}
+ \frac{1}{2} w_\lambda g^\mu{}_\alpha M^{\lambda \nu}
+ \frac{1}{2} w_\lambda g^\nu{}_\alpha M^{\mu \lambda}
\\&\quad
- \frac{1}{2} w^\mu M_\alpha{}^\nu
- \frac{1}{2} w^\nu M^\mu{}_\alpha
,
\\
\delta (g_{\mu\nu} \bR) &= -w^\lambda{}_\lambda
.
\ees
The terms with $w$ without a covariant derivative correspond to global
transformations. It is easy to verify that \nec{10.18a} remains invariant
under global longitudinal transformations by applying the simultaneous
rescaling $z\to \xi^{-1} z$ (which in turn implies $N_M \to \xi N_M$ or
$\delta N_M = w N_M$). Therefore it is only necessary to keep terms with
$w_\mu $ and $w_{\mu\nu}$ in the variation of \nec{10.18a}. After variation
and integration by parts in $x$ to remove terms with $w_{\mu\nu}$ (two
derivatives), one finds that the result cancels upon integration over
$p$. Once again we have checked this through third order in an expansion in
powers of $H^{\mu\nu}$.

The cancellation of transverse and longitudinal variations of the effective
action is a highly non trivial check of \Eq{10.18b}.

\section{\textsf{ Computation of the diagonal matrix elements
}
\label{sec:12}}

\subsection{\textsf{ Method of covariant symbols
}
\label{sec:12.1}}

The diagonal matrix elements can be computed using the method of covariant
symbols. This method was introduced in \cite{Pletnev:1998yu} for flat
spacetime and extended to curved spacetime in \cite{Salcedo:2006pv} where it
is described in great detail. It has also been extended to finite temperature
in \cite{Moral-Gamez:2011wcb,Brauner:2011vb}.  A summary can be found in
\cite{Garcia-Recio:2019iia} (Sec. 3.4 and App. B). Nevertheless, as the
method is not widely known, and to have a more self-contained work, we give
some details here.

For an operator $\hat\cO$, constructed with $\nabla_\mu$ plus some purely
multiplicative fields, such those in the basis \nec{8.6}, its covariant symbol
will be denoted $\overline{\cO}$. This quantity is obtained by applying two
successive similarity transformations\mfootnote{This formula is schematic, see
  full construction in \cite{Salcedo:2006pv}.}
\be
\overline{\cO}  := 
e^{-\frac{1}{2}\{\nabla_\mu,\partial^\mu\}} e^{-\xi^\alpha p_\alpha} \,
\hat{\cO} \,
e^{\xi^\beta p_\beta} e^{\frac{1}{2}\{\nabla_\nu,\partial^\nu\}}
\big|_{\xi^\mu=0}
\label{eq:12.1}
\,.
\ee
Here $\{,\}$ denotes anticommutator, $\xi^\mu$ are the Riemann coordinates
corresponding to the affine connection in $\nabla_\mu$ located at the point
$x$ where the diagonal matrix element $\esp{x|\hat{\cO}|x}$ will be
evaluated. $p_\mu$ is a momentum variable and
$\partial^\mu \equiv \partial/\partial p_\mu$.  For convenience, we use a
purely imaginary variable, $p_\mu=i k_\mu$, $k\in\R^d$ but $d^dp \equiv
d^dk$. This definition of covariant symbol holds for $\nabla_\mu$ having any
affine connections (e.g., with torsion); here we assume the Levi-Civita
connection for the coordinate indices, plus the gauge connection.

The original operator $\hat\cO$ acts only in $x$-space (and possibly in some
internal space) while its covariant
symbol $\overline{\cO}$ acts both on $x$- and $p$-spaces. The first remarkable
property of $\overline{\cO}$ is that this operator is {\em multiplicative with
  respect to $x$} (although not with respect to $p$). This means that it
commutes with functions of $x$ which are coordinate-scalar and gauge-singlet.

The second important property of the map $ \hat{\cO} \mapsto \overline{\cO}$
is that, being a similarity transformation, it is a faithful algebra
homomorphism (which also preserves hermiticity). Therefore, the covariant
symbol of a pseudo-differential operator of the type $f(\nabla_\mu, X)$ is
simply $f(\overline{\nabla}_\mu, \overline{X})$. This implies that it is
sufficient to obtain (once and for all) the covariant symbol of a few basic
blocks and once this is done there is no need to go back to the original
definition in \Eq{12.1}. For instance,
\bes
\overline{\nabla}_\mu &= p_{\mu }
- \frac{1}{4} \{ Z_{\alpha \mu },\partial^{\alpha } \}
+ \frac{1}{12} \{ [ Z_{\alpha \mu } , p_{\beta } ] , \partial^{\alpha }
\partial^{\beta }\}
+ O(p^{-2})
,
\\
\overline{X} &= X
- X_{\alpha} \partial^\alpha
+ \frac{1}{2!} X_{\alpha \beta } \partial^\alpha \partial^\beta
+ O(p^{-3})
,
\\
\overline{Z}_{\mu\nu} &= Z_{\mu\nu}
- \frac{1}{2} \{Z_{\alpha\mu\nu},\partial^\alpha\}
+ O(p^{-2})
.
\label{eq:8.13a}
\ees
Extensive formulas can be found in \cite{Salcedo:2006pv,Garcia-Recio:2019iia}.
As advertised, explicit $\nabla_\mu$ are no longer present and the covariant
symbol is a multiplicative operator.  The expressions take the form of an
expansion in powers of $\nabla/p$, i.e., with terms ordered by the number of covariant derivatives or
equivalently by powers of $p_\mu$ (counting $\partial^\mu$ as
$O(p^{-1})$). Here $X$ is any purely multiplicative operator (i.e., not
containing $\nabla_\mu$ nor $Z^R_I$), so in particular
\be
\overline{M}_{\mu\nu} = M_{\mu\nu}
- M_{\alpha \mu\nu} \partial^\alpha
+ \frac{1}{2!} M_{\alpha \beta \mu\nu} \partial^\alpha \partial^\beta
+ O(p^{-3})
.
\ee
And of course, $\overline{Z}_{\mu\nu} = [\overline{\nabla}_\mu,
  \overline{\nabla}_\nu ]$ is verified.

  It might seem that $Z_{\mu\nu}$ or $Z_{\alpha\mu\nu}$ would commute with $p_\lambda$ or
  $\partial^\lambda$ in \Eq{8.13a}, but in fact this is not
  so. Since $p_\mu$ and $\partial^\mu$ carry coordinate indices, \Eq{8.6a}
  applies and one has instead
\be
[Z_{\mu\nu}, p_\alpha] = R_{\mu\nu\alpha\beta}p_\beta,
\qquad
    [Z_{\mu\nu}, \partial^\alpha] = R_{\mu\nu\alpha\beta}\partial^\beta
    \,.
\label{eq:12.4}
    \ee    
The Eqs. in \nec{8.13a} are written so that the hermiticity properties are
manifest. Note that the metric does not appear in the covariant symbols of
$\nabla_\mu$ or $X$, in fact those formulas hold for a completely arbitrary
torsionless connection.\mfootnote{In this sense $\overline{\nabla}_\mu$ or
  $\overline{X}$ are truly {\em universal}.} The concrete connection will show
up through the action of $Z_{\mu\nu}$, as in \Eq{8.17}. In our case we have also
$\overline{g}_{\mu\nu} = g_{\mu\nu}$ since $\nabla_\mu$ is metric preserving.

The last important property of the covariant symbols to be noted is their
relation with diagonal matrix elements,
\be
\esp{x|\hat\cO|x} =
\frac{1}{\sqrt{g}} \int \frac{d^dp}{(2\pi)^d} \overline{\cO} \,\unit
.
  \label{eq:8.10}
  \ee
Here the object $\unit$ (which is usually not written explicitly) is the
function identically equal to 1 in momentum space. In practice this means that
the operator $\partial^\mu$ in $\overline{\cO}$ vanishes when it is placed on
the right. On the other hand $\partial^\mu$ is also zero when placed on left
due to the integral over $p_\mu$.

The quantity $\overline{\cO}\unit$ (with all the $\partial^\mu$ already
canceled by moving them to the right) is just a function of $x$ and $p$, and
is closer to the standard definition of the symbol of a pseudo-differential
operator. Hence the function $\overline{\cO}\unit$ (multiplicative with
respect to $p_\mu$) is what is needed to obtain the diagonal matrix elements
but the full covariant symbol $\overline{\cO}$ is the object carrying a
faithful algebra representation.

If instead of \Eq{12.1} only the first similarity transformation is applied,
$\hat{\cO} \mapsto e^{-\xi^\alpha p_\alpha} \, \hat{\cO} \, e^{\xi^\beta
  p_\beta}$, one obtains $f(\nabla,X)\mapsto f(\nabla+p,X)$, and the result is
the method of non-covariant symbols,
\be
\esp{x|f(\nabla,X)|x} =
\frac{1}{\sqrt{g}} \int \frac{d^dp}{(2\pi)^d} f(\nabla+p,X)
.
\label{eq:8.10c}
\ee
The integrand is not a multiplicative operator (nor manifestly covariant) but
it becomes so after integration over $p_\mu$. What the additional similarity
transformation achieves in \Eq{12.1} is precisely to have an integrand which
is manifestly covariant by systematically applying integration by parts in
$p$-space. Further details on the method of non-covariant symbols are given in
App. \ref{app:e}.

The manifest covariance of $\overline{\cO}$ (and hence $\overline{\cO}\unit$)
follows from using Riemann coordinates at $x$ (rather than the Synge function
as in the Schwinger-DeWitt approach), so the method of covariant symbols is
suited to computing diagonal matrix elements (within a derivative expansion
approach), but not for non-diagonal matrix elements. On the other hand the
method makes no assumptions on $\hat\cO$ so it works equally well even for a
pseudo-Laplacian like $\nabla^2_M$ with a ``metric'' which is non-Abelian.

We will illustrate the use of the method of covariant symbols below, but its
application is straightforward. $\overline{\cO}$ is obtained from $f(\nabla,X)
\mapsto f(\overline{\nabla},\overline{X})$, and $\overline{\nabla}_\mu$,
$\overline{X}$ are taken from the already compiled tables to the required
order. The quantity $\overline{\cO}$ so obtained contains only
purely multiplicative operators (covariant derivatives of $X$), plus
$Z_{\mu_1\cdots\mu_n}$, $p_\mu$ and $\partial^\mu$.

The natural next step is to remove all $\partial^\mu$ by moving them to the
right where they vanish, acting on all dependence on $p_\mu$. In doing so the
commutator between $Z_{\mu_1\cdots\mu_n}$ and $\partial^\mu$ has to be
applied. This generates some Riemann tensors. One can also choose to move
some of the $\partial^\mu$ to the left (where they also vanish) if this
produces a smaller number of terms. The two choices are related through
integration by parts in momentum space.

Since $p_\mu$ does not commute with $Z_{\mu_1\cdots\mu_n}$, in order to carry
out the momentum integration it will be convenient to move all the
$Z_{\mu_1\cdots\mu_n}$ together, to the right (or to the left) using their
commutation relations. This produces more coordinate curvatures and also gauge
curvatures $F_{\mu\nu}$. All these manipulations produce a diagonal matrix
element which does not assume a particular vector space for the action of the
operator $\hat{\cO}$. The concrete space is used when the operators
$Z_{\mu_1\cdots\mu_n}$ are removed after its action is evaluated, as in
\Eq{8.17}.

Once the $\partial^\mu$ have been eliminated and all the
$Z_{\mu_1\cdots\mu_n}$ are together (or eliminated) it only remains to carry
out the momentum integration, if possible.  When the non-polynomial dependence
on $p_\mu$ comes from a single (and Abelian) metric the integrals can often be
obtained explicitly.  If there are two metrics $g_{\mu\nu}$ and $M^{\mu\nu}$
(are even more so when the latter is non-Abelian) the integrals cannot be
evaluated in closed form in general. In this case one should be aware of
ambiguities in the final expression due to integration by parts in momentum
space. Complicated expressions can occasionally reach a simpler form through
integration by parts in $p_\mu$. For this reason the numbers quoted in Table
\ref{tab:1} are upper bounds.

An explicit application of the method of the covariant symbols to illustrate
the procedure just described is displayed in App. \ref{app:f}, by computing
one of the universal functional traces of \cite{Barvinsky:1985an}.

In what follows we proceed to give details of the calculation of $\Gammadiv_S$
and $\Gammadiv_{L,0}$.

\subsection{\textsf{ Calculation of $\Gammadiv_S$ in $d=2$
}
\label{sec:12.2}}

The starting point is \Eq{10.11a}. Applying the covariant symbols formula
\Eq{8.10},
\be
\esp{x|\hat\cO|x} = \esp{ \overline{\cO} }_p
\ee
(with $\esp{~}_p$ defined in \Eq{8.32c} and the $\unit$ is implicit) one
has\mfootnote{For a multiplicative $X$, $\esp{x|\cO X|x}=\esp{x|\cO |x} X(x)$,
  consistently
  $\esp{ \overline{\cO}\,\overline{X}}_p = \esp{\overline{\cO}}_p X(x)$ since
  $\overline{X} - X $ contains $\partial^\mu$ but not $p_\mu$ (see \Eq{8.13a}), and such terms
  vanish inside $\esp{~}_p$.}
%
\bes
    \Gamma_{S,1} &=
    \Bigg\langle
    \tr_1\Big(
    -\frac{1}{2} \overline{\cO}_{19}\, Y_{\mu\nu}
    \Big)
    \\&\quad
    + \tr_0\Big(
    -\frac{1}{2} \overline{\cO}_2 W
    -\frac{1}{2} (\overline{\cO}_1)_\mu P_\mu
    + \frac{1}{4} \overline{\cO}_2 \,P_{\mu\mu}
    + O(p^{-3})
    \Big)
    \Bigg\rangle_{x,p}
    \\
    \Gamma_{S,2} &=
    \\ &
    \Bigg\langle\tr_0\Big(
      \frac{1}{2} (\overline{\cO}_3)_{\mu\nu}
      \cM_{\mu\alpha} \cM_{\alpha\nu}
    -\frac{1}{4} (\overline{\cO_4[P_\mu]})_{\mu\nu} \, P_\nu
    + O(p^{-3})
    \Big)
    \Bigg\rangle_{x,p}
    \label{eq:8.33}
\ees
Since $\overline{\nabla}_\mu = O(p_\mu)$, while for multiplicative operators
$\overline{X} = O(1)$ for large $p_\mu$, the terms $O(\nabla^{-n})$ in
$\hat{\cO}$ become $O(p^{-n})$ in $\overline{\cO}$.

Within dimensional regularization $\esp{p^{-n}}_p$ vanishes for all $n$ with
the exception $n=d$. Hence we need to isolate terms $1/p^2$ in $d=2$, and in
particular more UV convergent terms can be neglected.

Of the basic operators present in the formula, $\cO_{2,3,4,19}$, are of
$O(\nabla^{-2})$. Therefore the covariant symbols of the latter only require
the leading terms of the building blocks:
%
\bes%
\overline{\nabla}_\mu
&%
= p_\mu + O(p^{-1}),
\qquad
\overline{\Deltag} = -N_g + O(p^{-3}),
\qquad
\\%
\oDeltaM
&%
= -N_M + O(p^{-3}),
\qquad
\overline{X} = X+O(p^{-1}),
\label{eq:12.9}
\ees%
where $X$ is any purely multiplicative operator. Here we have introduced the
definitions
\be
N_g \equiv (-g^{\mu\nu} p_\mu p_\nu)^{-1}
,\quad
N_M \equiv (-M^{\mu\nu} p_\mu p_\nu)^{-1}
.
\label{eq:8.14a}
\ee
Note that $N_M$ is a matrix in gauge space. This produces
%
\bes%
\overline{\cO}_2 &= -N_M + O(p^{-3}),
\qquad
\\
\overline{\cO}_3 &= N_M N_g p_\mu p_\nu + O(p^{-3}),
\qquad
\\%
\overline{\cO_4[X]}_{\mu\nu}
& = N_M X N_M p_\mu p_\nu + O(p^{-3}),
\qquad
\\
\overline{\cO}_{19} &= -N_g + O(p^{-3})
.
\ees%
The terms shown explicitly are homogeneous of degree $p^ {-2}$.

The remaining operator $\cO_1$ is $O(\nabla^{-1})$ and its covariant symbol
$O(p^{-1})$. To isolate the $1/p^2$ term we have to take one more term in the
expansion. The expansion in \nec{8.13a} is effectively in powers of
$\nabla/p$, so terms with one more covariant derivative are needed. Note
that in this counting $Z_{\mu_1,\cdots,\mu_n}$ counts as $O(\nabla^n)$. From
its definition $(\cO_1)_\lambda \equiv \DeltaM \nabla_\lambda$, one obtains
\be
(\overline{\cO}_1)_\lambda
=
  \oDeltaM \overline{\nabla}_\lambda
\ee
with
\be
\oDeltaM = (\overline{\nabla}_\mu \overline{M}_{\mu\nu} \overline{\nabla}_\nu
)^{-1}
.
\ee
The expansion of $\overline{\nabla}_\lambda$ in \Eq{12.9} is already
sufficient, but
$\oDeltaM$ needs to be expanded to order $1/p^3$ which in turn requires
$\overline{M}_{\mu\nu}$ to order $1/p$
\bes
\oDeltaM &= \left(
  p_\mu (M_{\mu\nu} - M_{\alpha \mu\nu} \partial^\alpha ) p_\nu
  + O(1)
\right)^{-1}
\\
&= -N_M + N_M p_\mu M_{\nu\mu\alpha} \partial^\nu p_\alpha N_M
+ O(p^{-4})
.
\label{eq:8.15}
\ees
Therefore,
\bes
(\overline{\cO}_1)_\lambda
&= -N_M p_\lambda + N_M p_\mu M_{\nu\mu\alpha} \partial^\nu p_\alpha  N_M p_\lambda
+ O(p^{-3})
.
\ees

To the order needed, the operators $\overline{\cO}_{2,3,4,19}$ do not have any
$\partial^\mu$ hence they already coincide with $\overline{\cO}_n\unit$. For
$\overline{\cO}_1$ the momentum derivatives can be moved to the right using
\be
[\partial^\mu, p_\nu] = \delta^\mu_\nu,
\qquad
[\partial^\mu, N_M ] = 2 p_\nu N_M M^{\mu\nu} N_M 
\ee
In this way\mfootnote{Actually for convenience here $\partial^\mu$ has been
  moved to the left (and then removed) so the RHS differs from
  $\overline{\cO}\unit$ \,by terms which vanish upon momentum integration.}
%
\bes%
&%
(\overline{\cO}_1)_\lambda \unit = 
- N_M p_\lambda
- N_M M_{\mu\mu\nu} N_M p_\lambda p_\nu
\\&\quad%
- 2 N_M M_{\mu\nu} N_M M_{\mu\alpha\beta} N_M p_\lambda p_\nu p_\alpha p_\beta
\\&\quad%
+ O(p^{-3})
\label{eq:8.14}
\ees%
The first term is homogeneous of degree $1/p$ and the two other explicit terms
are homogeneous of degree $1/p^2$. The first term vanishes within momentum
integration due to parity. For the same reason all odd order terms have been
omitted in Table \ref{tab:1}. In any case only the contributions $1/p^d$ are
relevant in $d$ dimensions for $\Gammadiv_1$.

No operators $Z_I$ appear in the expansions of $\overline{\cO}_{1,2,3,4,19}$
to the order required in \Eq{8.33}. The most UV divergent operator is
$\cO_1=O(\nabla^{-1})$, so its leading term is $1/p$ and the term relevant in
$d=2$, $1/p^2$, comes from contributions of the type $\nabla/p^2$, while $Z_I$
needs at least two covariant derivatives.

Operators $Z_I$ do appear in $d=4$. Just as $\partial^\mu$, the operators
$Z^R_I$ have to be resolved before having a useful expression for
$\esp{x|\cO_n|x}$. To this end, the operators $Z^R_I$ can be moved to the left
or to the right, using their commutation relations noted in \nec{8.6a},
including \nec{12.4}. In accordance with the choice in \Eq{8.4}, where the
multiplicative operators $A_n$ have been placed at the right, $Z^R_I$ should
be moved to the left. This allows to apply the rules in \Eq{8.17} (see
App. \ref{app:a}).

It remains to insert in \Eq{8.33} the various expressions for
$\overline{\cO}_n\unit$ just obtained .  In the resulting expression, the
terms involving $P_\mu$ in $\Gamma_{S,1}$ are not manifestly Hermitian. This
can be fixed by applying integration by parts in $p$ and $x$
spaces\mfootnote{As shown in App. C of \cite{Garcia-Recio:2019iia}, $p_\mu$
  can be treated as a constant when integrating by parts in $x$-space, when
  $\nabla_\mu$ and $Z^R_I$ are no longer present.} as well as the trace cyclic
property. This gives,
%
\bes
\Gamma_{S,1}^\div &=
\Big\langle 
 \frac{1}{2} N_g Y_{\mu\mu}
+ \frac{1}{2} N_M W
 - \frac{1}{4} N_M P_{\mu\mu}
\\&\qquad
+ \frac{1}{2} N_M M_{\mu\nu} N_M M_{\mu\alpha\beta} N_M P_\lambda
p_\nu p_\alpha p_\beta p_\lambda
\\&\qquad
- \frac{1}{2} N_M M_{\mu\alpha\beta} N_M M_{\mu\nu} N_M P_\lambda
p_\nu p_\alpha p_\beta p_\lambda
\Big\rangle_{x,\,p,\, g}
,
\\
\Gamma_{S,2}^\div &=
\Esp{
\frac{1}{2} N_g N_M \cM_{\mu\alpha}\cM_{\alpha\nu} \, p_\mu p_\nu
- \frac{1}{4} N_M P_\mu N_M P_\nu p_\mu p_\nu
}_{x,\,p,\, g}
\,.
\label{eq:8.36a}
\ees
The integrand is a homogeneous function of $p$ of degree $-2$. It only remains
to extract the $1/\epsilon$ coefficient to isolate the UV divergent
contributions.  The details are given below, in Sec. \ref{sec:8.c}. An
application of the rules provided there immediately produces the result
quoted in \Eq{8.36b}.

In \Eq{8.14} there is one term of degree $p^{-1}$ and two terms of degree
$p^{-2}$. Table \ref{tab:1} shows the number of terms of each degree for the
diagonal matrix elements of the operators $\cO_n$. Only even orders are
displayed since odd orders vanish upon integration over $p_\mu$ in any parity
preserving regularization, such as dimensional regularization. For a given
operator, the number of terms increases rapidly with (minus) the
degree. Nevertheless the number of terms displayed in the table is an upper
bound, this number is subject to
variations due to various identities which allow to write a given expression in
different forms.  Such identities include integration by parts in momentum
space and reordering of
the covariant derivatives acting on a tensor due to the Jacobi identity,
 \be
    [\nabla_\mu,[\nabla_\nu, X]] =
    [\nabla_\nu,[\nabla_\mu, X]]
+ [Z_{\mu\nu},X] 
.
\label{eq:8.19}
 \ee
 Furthermore, integration by parts in $x$-space and trace cyclic property is
 allowed within the functional trace operations $\Tr_{0,1}$ in \Eq{8.10a}. We
 have not attempted a systematic minimization of the number of terms as there
 is no practical procedure to do this, and in any case we do not expect a
 significant reduction in the length of the expressions. An exception is the
 operator $\cO_2$ at $p^{-4}$ which is used below, \Eq{10.18}, in the
 computation of $\Gamma_{L,0}$, in Sec. \ref{sec:10.C}.

 Nevertheless, it should be noted that recently important progress has been
 achieved in the counting and classification of allowed independent terms in
 effective field theories, through the construction of Hilbert series of the
 operator basis \cite{Henning:2017fpj}. In this technique the basic blocks
 (fields or composite operators) plus their symmetrized derivatives are
 identified with representations of the $d$-dimensional conformal
 group.\mfootnote{Alternatively, cohomological techniques can be applied
   \cite{Henning:2017fpj}.} Computation of the Clebsch-Gordan series then
 allows to obtain generating functions for basis operators and count them. The
 key point is that both, equation of motion as well as integration by part
 identities, are automatically accounted for, in addition to spacetime and
 internal group symmetries. The method has been successfully applied to pure
 Einstein relativity and also to general relativity combined with the Standard
 Model of particle physics \cite{Ruhdorfer:2019qmk}. The adaptation of the
 Hilbert series technique to obtain basis of operators in diagonal matrix
 elements and the effective action contributions as those displayed in
 \Eq{8.36a} or \Eq{10.18a} would be extremely interesting, and more so in
 $d=4$ where the number of terms becomes huge. Serious complications arise due
 to the presence of an additional momentum variable, with its own integration
 by parts identities, and also the existence of constraints relating some of
 the building blocks, such as $M^{\mu\nu}$ and $N_M$. No attempt will be made
 here to adapt the promising technique of Hilbert series to working with
 covariant symbols, we defer such a study to future work.

\subsection{\textsf{ Calculation of $\Gammadiv_{L,0}$ in $d=2$
}
\label{sec:12.3}}

Due to the similarity between the operators $\cO'_2$ and $\cO_2$, the
expression of their diagonal matrix elements are identical when written in
terms of $N_M$, with the only proviso of using the new definition
\be
N_M = (-z-M^{\mu\nu} p_\mu p_\nu )^{-1}
\ee
for $\cO'_2$, instead of that in \nec{8.14a}.

The UV divergent terms in $\Gamma_{L,0}$ must have exactly $d$ covariant
derivatives. The operator $\cO_2'$ is of UV degree $O(\nabla^{-2})$, so the
leading term in $\overline{\cO'}_2$ is $O(p^{-2})$ and no derivatives.
Since one still needs to expand its covariant symbol to $d$ covariant
derivatives, the order $p^{-d-2}\nabla^d$ is needed. The proper divergence $O(p^{-d})$ is only
recovered after integration over $z$, a parameter of dimension squared mass.

We present results only for the case $d=2$. For the terms of order
exactly $p^{-4}$, the calculation gives the following result
%
\bes
&\esp{x|\cO'_2|x}^{(-4)} =
\\&\quad
\Big\langle
 4 {N_M} \, M_{\mu  \nu } \, {N_M} \, M_{\mu  \alpha  \beta } \, {N_M} \, M_{\rho  \sigma  \lambda } \, {N_M} \, M_{\rho  \eta } \, {N_M} \, p_{\nu } \, p_{\alpha } \, p_{\beta} \, p_{\sigma } \, p_{\lambda } \, p_{\eta }
\\&\quad
+ 2 {N_M} \, M_{\mu  \nu } \, {N_M} \, M_{\mu  \alpha  \beta } \, {N_M} \, M_{\rho  \rho  \sigma } \, {N_M} \, p_{\nu } \, p_{\alpha } \, p_{\beta } \, p_{\sigma }
   \\&\quad
+ 2 {N_M} \, M_{\mu  \mu  \nu } \, {N_M} \, M_{\alpha  \beta  \rho } \, {N_M} \, M_{\alpha  \sigma } \, {N_M} \, p_{\nu } \, p_{\beta } \, p_{\rho } \, p_{\sigma }
   \\&\quad
+{N_M} \, M_{\mu  \nu } \, {N_M} \, M_{\mu  \alpha  \beta  \rho} \, {N_M} \,
M_{\alpha  \sigma } \, {N_M} \, p_{\nu } \, p_{\beta } \, p_{\rho} \,
p_{\sigma } 
\\&\quad
+{N_M} \, M_{\mu  \nu} \, {N_M} \, M_{\alpha  \mu  \beta  \rho } \, {N_M} \,
M_{\alpha  \sigma } \, {N_M} \, p_{\nu } \, p_{\beta } \, p_{\rho } \, p_{\sigma }
\\&\quad
+\frac{2}{3} {N_M} \, M_{\mu  \nu } \, {N_M} \, M_{\alpha  \beta } \, {N_M} \,
M_{\rho  \sigma } \, {N_M} \, {R}_{\mu  \alpha  \rho  \lambda } \, p_{\nu} \,
p_{\beta } \, p_{\sigma } \, p_{\lambda } 
\\&\quad
-\frac{2}{3} {N_M} \, M_{\mu  \nu } \, {N_M} \, M_{\alpha  \beta } \, {N_M} \,
M_{\rho  \sigma } \, {N_M} \, {R}_{\mu  \lambda  \alpha  \rho } \, p_{\nu } \,
p_{\beta } \, p_{\sigma} \, p_{\lambda }
   \\&\quad
+{N_M} \, M_{\mu  \mu  \nu } \, {N_M} \, M_{\alpha  \alpha  \beta } \, {N_M}
\, p_{\nu } \, p_{\beta } 
\\&\quad
-{N_M} \, M_{\mu  \nu } \, F_{\mu  \alpha } \, {N_M} \, M_{\alpha  \beta} \,
{N_M} \, p_{\nu } \, p_{\beta } 
\\&\quad
-{N_M} \, M_{\mu  \nu } \, {N_M} \, F_{\mu  \alpha } \, M_{\alpha  \beta } \,
{N_M} \, p_{\nu } \, p_{\beta} 
   \\&\quad
   -\frac{1}{6} {N_M} \, M_{\mu  \nu } \, {N_M} \, {\cR}_{\mu  \nu }
   \Big\rangle_p
   .
\label{eq:10.18}
\ees
%
The expression obtained after applying the method of covariant symbols has
been simplified by using integration by parts in momentum space, also
achieving a manifest Hermitian form (since $\cO_2$ is Hermitian).

It can be noted that the expression in \nec{10.18} holds for arbitrary $d$.
For $d=2$ one can use the identities in \nec{10.19}.  Also we integrate by
parts with respect to $x$ to have at most one covariant derivative on
$M_{\mu\nu}$. This gives for the effective action
%
\bes
&\Gamma_{L,0}^\div
=
\frac{1}{2}\Big\langle
\\&\quad
-2  N_M \,M_{\mu  \nu }\, N_M \,M_{\mu  \alpha  \beta
   }\, N_M \,M_{\rho  \sigma }\, N_M \,M_{\rho  \lambda  \eta }\, N_M \,p_{\nu
}\,p_{\alpha }\,p_{\beta }\,p_{\sigma }\,p_{\lambda }\,p_{\eta }
\\&\quad
-2  N_M \,M_{\mu  \nu
    \alpha }\, N_M \,M_{\mu  \beta }\, N_M \,M_{\rho  \sigma  \lambda
   }\, N_M \,M_{\rho  \eta }\, N_M \,p_{\nu }\,p_{\alpha }\,p_{\beta }\,p_{\sigma
}\,p_{\lambda }\,p_{\eta }
\\&\quad
- N_M \,M_{\mu  \nu }\, N_M \,M_{\mu  \alpha  \beta
   }\, N_M \,M_{\alpha  \rho  \sigma }\, N_M \,p_{\nu }\,p_{\beta }\,p_{\rho }\,p_{\sigma
}
\\&\quad
+ N_M \,M_{\mu  \nu }\, N_M \,M_{\alpha  \alpha  \beta }\, N_M \,M_{\mu  \rho
   \sigma }\, N_M \,p_{\nu }\,p_{\beta }\,p_{\rho }\,p_{\sigma }
\\&\quad
- N_M \,M_{\mu  \nu
    \alpha }\, N_M \,M_{\beta  \mu  \rho }\, N_M \,M_{\beta  \sigma }\, N_M \,p_{\nu
}\,p_{\alpha }\,p_{\rho }\,p_{\sigma }
\\&\quad
+ N_M \,M_{\mu  \nu  \alpha }\, N_M \,M_{\beta
    \beta  \rho }\, N_M \,M_{\mu  \sigma }\, N_M \,p_{\nu }\,p_{\alpha }\,p_{\rho
}\,p_{\sigma }
\\&\quad
+ N_M \,M_{\mu  \nu }\, N_M \,M_{\alpha  \beta
   }\, N_M \,M_{\mu  \rho }\, N_M \,\bR \,p_{\nu }\,p_{\alpha }\,p_{\beta }\,p_{\rho
}
\\&\quad
+ N_M \,M_{\mu  \mu  \nu }\, N_M \,M_{\alpha  \alpha  \beta
}\, N_M \,p_{\nu }\,p_{\beta }
\\&\quad
- N_M \,M_{\mu  \nu }\,F_{\mu  \alpha }\, N_M \,M_{\alpha  \beta }\, N_M \,p_{\nu
}\,p_{\beta }
\\&\quad
- N_M \,M_{\mu  \nu }\, N_M \,F_{\mu  \alpha }\,M_{\alpha  \beta
}\, N_M \,p_{\nu }\,p_{\beta }
\\&\quad
+\frac{2}{3}  N_M \,M_{\mu  \nu }\, N_M \,M_{\mu  \alpha }\, N_M \,\bR\,p_{\nu
}\,p_{\alpha }
\\&\quad
-\frac{1}{12}  N_M \,M_{\mu  \mu }\, N_M \,\bR
\Big\rangle_{x,p,g,z}
\label{eq:10.18a}
\ees
%
One could apply further the trace cyclic property, and also the identity $N_M
M_{\mu\nu} N_M p_\mu p_\nu = -N_M-z N_M^2$ in one of the terms, but no
simplification would be achieved. The final step is to extract the
$1/\epsilon$ coefficient from the radial part of the momentum integral, as
described in next subsection. This procedure yields the expression quoted in
\Eq{10.18b}.

\subsection{\textsf{Extraction of the UV divergent component}
\label{sec:8.c}}

Let us consider first the case when the parameter $z$ is not present, as in
\Eq{8.36a}.  Once the
operators $\partial_\mu$ and $Z^R_I$ have been removed, the structure of a
general term $\overline{\cO}_n$ to be integrated over $p_\mu$ is a sum of products with factors $N_M$, $N_g$, $p_\mu$, as well
as $p_\mu$-independent multiplicative operators $M_I,R_I,F_I$, etc. Hence the
momentum integral affects only terms of the form
\be
N_g^n (N_M)_{a_1 b_1} \cdots (N_M)_{a_m b_m} p_{\mu_1}\cdots p_{\mu_{2j}}
\equiv N_g^n N_M^{\otimes m} p^{\otimes 2j}
.
\ee
This monomial is homogeneous in $p_\mu$ with degree $2(j-n-m)$.

We are only interested in the UV divergent part and to extract it we will use
dimensional regularization, namely, in $d+2\epsilon$ dimensions with
$\epsilon\to 0^-$. For the UV divergent part, $d$ can be used instead of
$d+2\epsilon$ in the UV-finite contributions. Within dimensional
regularization, the integral
\be
I_d^{m,j} = \frac{1}{\sqrt{g}} \int \frac{d^{d+2\epsilon}p}{(2\pi)^d}
N_g^n N_M^{\otimes m} p^{\otimes 2j},
\ee
vanishes unless $d=2(n+m-j)$, which is assumed in what follows.

Recalling that $p_\mu=ik_\mu$ and $d^dp_\mu\equiv d^dk_\mu$,
\be
I_d^{m,j} = (-1)^j \frac{1}{\sqrt{g}} \int \frac{d^{d+2\epsilon}k_\mu}{(2\pi)^d}
N_g^n N_M^{\otimes m} k^{\otimes 2j}
.
\ee
Let us introduce a standard tetrad field $e_\mu^A(x)$,
\be
g_{\mu\nu} = e^A_\mu e^B_\nu \delta_{AB},
\qquad
e^A_\mu e^\mu_B = \delta^A_B
\ee
in such a way that
%
\bes%
k_\mu
&%
= e^A_\mu k_A, \qquad
d^dk_\mu = \sqrt{g} d^dk_A
,\qquad
\\
N_g
&%
= \frac{1}{k_A^2} , \qquad
N_M = \frac{1}{k_Ak_B M^{AB}}
\ees
thus
%
\bes%
(I_d^{m,j})_{\mu_1\cdots\mu_{2j}}
&%
=(I_d^{m,j})_{A_1\cdots A_{2j}}
e^{A_1}_{\mu_1} \cdots e^{A_{2j}}_{\mu_{2j}}
,
\qquad
\\%
(I_d^{m,j})_{A_1\cdots A_{2j}}
&%
= (-1)^j \int \frac{d^{d+2\epsilon}k_A}{(2\pi)^d}
\, N_g^n \,  N_M^{\otimes m} \, k_A^{\otimes 2j}
.
\ees%
The UV divergence comes from the radial part of the integral, hence we
introduce spherical coordinates,
\be
k_A = k \hk_A,  \qquad k = \sqrt{k_A^2}=N_g^{-1/2}, \qquad \hk_A^2=1
.
\ee
This allows to separate the integral into radial and angular-average factors,
\be
I_d^{m,j}
=
\frac{(-1)^j}{(4\pi)^{d/2}} \frac{2}{\Gamma(d/2)} 
I_\epsilon \hat{I}_d^{\,m,j}
,
\ee
where the angular average is
%
\bes%
\hat{I}_d^{\,m,j}
&%
=
\frac{\Gamma(d/2)}{2\pi^{d/2}}
\int d^{d-1}\Omega_{\hk}
\, \hat{N}_M^{\otimes m} \, \hk_A^{\otimes 2j}
\equiv
\esp{ \hat{N}_M^{\otimes m} \, \hk_A^{\otimes 2j} }_\ang
,
 \qquad
\\%
\hat{N}_M
&%
\equiv \frac{1}{\hk_A \hk_B M^{AB}}
 = N_M/N_g,
 \label{eq:8.27}
\ees%
and $I_\epsilon$ is the radial part (introducing a cutoff mass $m_0$ to avoid
a trivial infrared divergence for negative $\epsilon$), and using the
condition $2(n+m-j) = d$,
\be
I_\epsilon = \int_{m_0}^\infty dk k^{2\epsilon -1} =
-\frac{m_0^{2\epsilon}}{2\epsilon} = 
-\frac{1}{2\epsilon}+ O(1)
.
\ee
In summary, for the UV divergent part, one obtains
\be
I_d^{m,j,\div}
=
\frac{1}{(4\pi)^{d/2}\Gamma(d/2)} \frac{1}{\epsilon} (-1)^{j+1}
\, \hat{I}_d^{\,m,j}
.
\label{eq:8.29}
\ee
As noted, the angular averages $\hat{I}_d^{\,m,j}$ are perfectly UV convergent
and well-defined, but they cannot be written in closed form in general.

The analysis is similar for $\Gamma_{L,0}$, which involves an additional
integration over $z$. As mentioned, in this case and for $d=2$, the relevant terms
are of order $p^{-4}$, and to extract the UV divergent part one must integrate
over $z$ and $p_\mu$. The point can be elucidated following the steps shown
above (for $d=2$), noting that now $N_M$ contains $z$:
%
\bes
&\Esp{ p^{\otimes 2j} N_M^{\otimes (2+j)} }_{z,p}
=
\frac{1}{\sqrt{g}}
\int\frac{d^{2+2\epsilon}p}{(2\pi)^2}\int_\gamma \frac{dz}{2\pi i} \,
\\&\quad\times
\log(z)
\,p^{\otimes 2j}
\left(\frac{1}{-z-M^{\mu\nu}p_\mu p_\nu }\right)^{\otimes(2+j)}
\\&=
(-1)^j \int_{m_0}^\infty dk \,k^{2j+1+2\epsilon}
\int\frac{d\Omega}{(2\pi)^2}\int_\gamma \frac{dz}{2\pi i} \,
\\&\quad\times
\log(z)
 \,\hk^{\otimes 2j}
\left(\frac{1}{-z + k^2 M^{\mu\nu}\hk_\mu \hk_\nu }\right)^{\otimes(2+j)}
\ees
Applying the rescaling $z\to zk^2$, and noting that the induced term with
$\log(k^2)$ vanishes since no singularities are enclosed by the path $\gamma$
in the $z$ complex plane,
\bes
&\Esp{ p^{\otimes 2j} N_M^{\otimes (2+j)} }_{z,p}
= 
(-1)^{j+1} \frac{m_0^{2\epsilon}}{2\epsilon}
\int\frac{d\Omega}{(2\pi)^2}\int_\gamma \frac{dz}{2\pi i} \,
\\&\quad\times
\log(z)
\, \hk^{\otimes 2j}
\left(\frac{1}{-z + M^{\mu\nu}\hk_\mu \hk_\nu }\right)^{\otimes (2+j)}
\\&\quad=
\frac{(-1)^{j+1}}{4\pi\epsilon}
\esp{
  \hk^{\otimes 2j} \,
\hat{N}_M^{\otimes(2+j)}
}_{\ang,z}
+ O(1)
,
\label{eq:10.22}
\ees
where $\hat{N}_M=(-z + M^{\mu\nu}\hk_\mu \hk_\nu )^{-1}$. Applying this
angular average in \Eq{10.18a} yields \Eq{10.18b}.

\section{\textsf{Summary and conclusions}
\label{sec:sum}}

In this work we have addressed the problem of quantizing a system of, in
general, non-Abelian vector fields with a completely general local non-minimal
mass term coupling all of them. The case of $N$ Abelian fields is a particular
instance in our formulation. We make use of a background field approach.  A
remarkable result is that, although the mass term breaks gauge invariance, the
effective action is fully gauge (as well as coordinate) invariant beyond
tree-level, \Eq{3.5}. The technical problems present in the original theory
(namely, the UV region is blind to the mass term and so requires some type of
gauge-fixing) are satisfactorily removed by introducing a non-Abelian
Stueckelberg field, Eqs. \nec{5.1} and \nec{5.4}. This is done after
background and fluctuation fields have been separated and the Stueckelberg
field only affects the latter.  This auxiliary field is introduced linearly so
that the loop structure of the original theory is preserved. As a consequence
the computation of the UV divergent part of the effective action to one-loop
can be carried out systematically preserving coordinate and gauge symmetries
during the calculation. To this end we apply dimensional regularization and
the method of covariant symbols. This produces terms which are local, i.e.,
they contain a finite number derivatives of the external fields (no more that
$d$ in $d$ spacetime dimensions), however they are not polynomial with respect
to the mass term, a fact already observed in the simpler case of a single
vector field.  The one-loop effective action is expressed in \Eq{4.28} as a
sum of four terms, $\Gamma_\gh$ defined in \nec{6.8}, $\Gamma_{L,1}$ and
$\Gamma_{L,0}$ in \nec{7.7}, and $\Gamma_S$ in \neq{7.8a} (and expanded in \Eq{8.1}). The formalism is
developed for arbitrary spacetime dimensions and we present explicit results
for the two-dimensional case. Regrettably in the four-dimensional case too
many terms are produced (even selecting particular settings, such as purely
Abelian, or flat spacetime, or perturbative expansions) so their explicit
expression would be of little practical use. The explicit two-dimensional
results for the UV divergent component are displayed in \nec{9.4} for
$\Gamma_\gh$ and $\Gamma_{L,1}$, in \nec{8.36b} for $\Gamma_S$, and in
\nec{10.18b} for $\Gamma_{L,0}$.  Checks have been applied to the results of
the calculation.  Particularly stringent are the tests related to invariance
with respect to metric transformations in Sec. \ref{sec:10.D}. Perturbative
results are also presented in App. \ref{app:d}. Although not explicitly
discussed in the text, we have repeated most of the calculations of diagonal
matrix elements using the method of non-covariant symbols
\cite{Salcedo:2006pv,Moral-Gamez:2011wcb} to find an identical result, modulo
integration by parts in momentum space. In some cases the latter method has
produced shorter expressions. Details of the non-covariant method are
discussed in App. \ref{app:e}.

As already pointed out, the non-minimal mass-like
coupling discussed in this work (or also its Abelian version), respects
locality but introduces terms in the effective action which are non-polynomial
in the field $M^{\mu\nu}(x)$. All current efforts for an effective field
theory description of general relativity or the standard model of particles,
or both (e.g. \cite{Ruhdorfer:2019qmk}) naturally assume an expansion in local
and polynomial operators over some power of the cutoff (a new physics scale),
$\cO_\alpha(x)/\Lambda^n$, consistently with the renormalization group
analysis of Wilson \cite{Wilson:1974mb}. In this light, the analyses presented
in \cite{Ruf:2018vzq,Garcia-Recio:2019iia} and in this work should indicate
that a non-minimal coupling of the type \Eq{2.1} can be ruled out in vector
field theories, also in the non-Abelian setting. If the presence of such
non-minimal coupling could not be prevented through some mechanism (such as
the requirement of strict gauge invariance) one would be impelled to assume a
much larger class of effective field theories, including local but
non-polynomial operators. On the other hand, even in that case re-expansions
as that in App. \ref{app:d} would bring the expression again to the standard
form, requiring only local and polynomial composite operators, provided that
the scale $m$ can be interpreted as a proper cutoff of the theory and a
separation of the type $M^{\mu\nu}= m^2 g^{\mu\nu}+H^{\mu\nu}$ is somehow
natural.

The fluctuation operator in \Eq{4.15} is a rather involved one, due to the
presence of a non-Abelian field $M^{\mu\nu}$ coupling like a metric in the
$\phi$-sector. Chan's method or even the Schwinger-DeWitt technique are not
readily available to deal with such term. Yet the formalism of covariant
symbols could be applied also to $\Gamma_{L,0}$, upon introduction of a
parametric form to remove the logarithm. In fact the method of covariant
symbols is a practical and easy-to-use tool to obtain diagonal matrix elements
of local operators $f(\nabla,X)$ provided the dependence on $\nabla$ is of
rational type. This latter requirement follows from the fact that, in
practice, the covariant symbols are only obtained as an expansion in powers of
$\nabla/p$. It can also be noted that the covariant symbols depend only on the
connection, the presence of a metric is not required, as the Riemann
coordinates can be defined directly from the connection
\cite{Gusynin:1990bu}. The method can be used to obtain not only the
counterterms but also covariant derivative expansions of the effective action
itself \cite{Salcedo:2000hx,Salcedo:2007bt,Garcia-Recio:2009jso}. In
particular, the treatment of fermionic modes in curved spacetime poses no
special problems once the spin connection is included in the covariant
derivative \cite{Motilla}. The extension of the method for finite temperature
also exists \cite{Moral-Gamez:2011wcb,Brauner:2011vb} (but not yet for
temperature and curvature at the same time). The method of covariant symbols
should apply whenever the generalized Schwinger-DeWitt technique applies, so
it can be used as an alternative approach in the analysis quantum field
theories in curved spacetime, or effective field theories involving gravity,
or in the study of the newly developed Proca theories noted in the
Introduction.

\acknowledgments
I thank C. Garcia-Recio for suggestions on the manuscript and A.O. Barvinsky
for critical remarks.
This work has been partially supported by
MCIN/AEI/10.13039/501100011033 under grant PID2020-114767 GB-I00,
by the Junta de Andaluc{\'\i}a (grant No. FQM-225),
by the FEDER/Junta de Andalucía-Consejería de Economía y Conocimiento
   2014-2020 Operational Program under grant A-FQM-178-UGR18,
and by the Consejer{\'\i}a de Conocimiento,
Investigaci\'on y Universidad, Junta de Andaluc{\'\i}a and European Regional
Development Fund (ERDF), ref. SOMM17/6105/UGR.

\appendix

\section{\textsf{Derivation of some formulas}
  \label{app:b}}

\subsection{\textsf{Derivation of \Eq{2.20}}}

Applying integration by parts in the first term in \neq{2.20a} (using the
notation of \nec{8.32})
\bes
&
\frac{1}{4}\Esp{ (\cA_{\mu\nu}^a-\cA_{\nu\mu}^a)^2}_x
\\&=
\frac{1}{2} \Esp{  (\cA_{\mu\nu}^a)^2 -\cA_{\mu\nu}^a\cA_{\nu\mu}^a}_x
=
\frac{1}{2} \Esp{  (\cA_{\mu\nu}^a)^2 + \cA_{\nu}^a\cA_{\mu\nu\mu}^a}_x
\\ &=
\frac{1}{2} \Esp{  (\cA_{\mu\nu}^a)^2 + \cA_{\nu}^a\cA_{\nu\mu\mu}^a
+ \cA_{\nu}^a Z_{\mu\nu}^{ab} \cA_{\mu}^b }_x
\\& =
\frac{1}{2} \Esp{  (\cA_{\mu\nu}^a)^2 - \cA_{\nu\nu}^a\cA_{\mu\mu}^a
  + \cA_{\nu}^a (    F_{\mu\nu}^{ab} \cA_{\mu}^b
   + R_{\mu\nu\mu\alpha} \cA_{\alpha}^a) }_x
\\& =
\frac{1}{2} \Esp{  (\cA_{\mu\nu}^a)^2 - (\cA_{\mu\mu}^a)^2
  - F_{\mu\nu}^{ab} \cA_{\mu}^a \cA_{\nu}^b
  + \cR_{\mu\alpha} \cA_{\alpha}^a) }_x
.
\ees
Added to the other term in \neq{2.20a}, $\esp{-\frac{1}{2} F_{\mu\nu}^{ab}
  \cA_{\mu}^a \cA_{\nu}^b}$, produces \neq{2.20}.

\subsection{\textsf{Derivation of \Eq{2.14}}}

From the change of variables
\be
\cA_\mu^a = \cB_\mu^a + \phi_\mu^a,
\qquad
\chi^a = \cB^a_{\mu\mu}
\ee
one obtains
\be
\frac{\partial(\cA^a_\mu,\chi^a)}{\partial(\cB^b_\nu,\phi^b)}
=
\begin{pmatrix}
  \delta_{ab} g_\mu{}^\nu & \delta_{ab}\nabla_\mu
  \\
  \delta_{ab}\nabla^\nu & 0
\end{pmatrix}
\ee
(with rows for the numerator and columns for the denominator). More
conveniently, doing the change of variables in two steps $(\cB,\phi) \to
(\cA,\phi) \to (\cA,\chi)$,
\be
\frac{\partial(\cA^a_\mu,\chi^a)}{\partial(\cB^b_\nu,\phi^b)}
=
\frac{\partial(\cA^a_\mu,\chi^a)}{\partial(\cA^c_\lambda,\phi^c)}
\frac{\partial(\cA^c_\lambda,\phi^c)}{\partial(\cB^b_\nu,\phi^b)}
\ee
using $\chi^a=\cA_{\mu\mu}^a- \phi^a_{\mu\mu}$ in the first factor, produces
\be
\begin{pmatrix}
  \delta_{ab} g_\mu{}^\nu & \delta_{ab}\nabla_\mu \\
  \delta_{ab}\nabla^\nu & 0
  \end{pmatrix}
=
\begin{pmatrix}
  \delta_{ac} g_\mu{}^\lambda & 0 \\
  \delta_{ac}\nabla^\lambda & -\delta_{ac} \nabla^2 
  \end{pmatrix}
\begin{pmatrix}
  \delta_{cb} g_\lambda{}^\nu & \delta_{cb}\nabla_\lambda \\
  0 & \delta_{cb}
\end{pmatrix}
.
\ee
The second matrix has unit determinant and likewise for the upper-left block
in the first matrix. This produces \Eq{2.14}.

\subsection{\textsf{Derivation of \Eq{2.22}}}

Starting from \nec{2.20a}, and using $\cA_\mu^a = \cB_\mu^a + \phi^a_\mu$, one
obtains terms of the types $\cB\cB$, $\cB\phi$ and $\phi\phi$. The terms
$\cB\cB$ are just those in \nec{2.20} with $\cA_\mu^a \to \cB_\mu^a$.

The terms $\cB\phi$ are given by twice \nec{2.20a} replacing one of the $\cA$ with 
$\cA^a_\mu\to\cB^a_\mu$ and the other one with
$\cA^a_\mu\to\phi^a_\mu$. This produces
\be
\left(S^{(2)}_{\rm kin}\right)_{\cB\phi} = \Esp{
  \frac{1}{2} (\phi_{\mu\nu}^a-\phi_{\nu\mu}^a)(\cB_{\mu\nu}^a-\cB_{\nu\mu}^a)
- F^{ab}_{\mu\nu} \phi^a_\mu \cB^b_\nu 
}_x
\ee
Using $\phi_{\mu\nu}^a-\phi_{\nu\mu}^a = F^{ab}_{\mu\nu}\phi^b$ and
integration by parts in the other term,
\bes
\left(S^{(2)}_{\rm kin}\right)_{\cB\phi} &= \Esp{
  - \phi^a F^{ab}_{\mu\nu}\cB_{\mu\nu}^b
  + F^{ab}_{\mu\mu\nu} \phi^a \cB^b_\nu
+
   F^{ab}_{\mu\nu} \phi^a \cB^b_{\mu\nu}
}_x
\\
&= \Esp{
     F^{ab}_{\mu\mu\nu} \phi^a \cB^b_\nu
}_x
\ees

Finally, the term $\phi\phi$ is half the previous one after the replacement
$\cB^a_\mu\to\phi^a_\mu$. This produces \neq{2.22}.

\section{\textsf{The operators $Z_{\mu_1\cdots\mu_n}$}
  \label{app:a}}

\subsection{\textsf{Definition and properties of $Z^R_{\mu_1\cdots\mu_n}$}}

The operator $Z_{\mu\nu}$ is defined as
\be
Z_{\mu\nu} = [\nabla_\mu,\nabla_\nu] = Z^R_{\mu\nu} + F_{\mu\nu}
\ee
$Z^R_{\mu\nu}$ acts on coordinate indices and $F_{\mu\nu}$ on gauge indices.
$Z^R_{\mu\nu}$ is multiplicative but not ``purely multiplicative'' (by
definition) as it is not diagonal in the coordinate indices. The higher rank
tensors are defined recursively, namely, (recall that $I=\mu_1\cdots\mu_n$
stands for a string of coordinate indices)
\be
Z_{\alpha I} = [\nabla_\alpha,Z_I] + \frac{1}{2}
\{\nabla_\lambda,R_{I\alpha\lambda} \}
\,.
\label{eq:B2}
\ee
The extra term ensures that $Z_I$ is a multiplicative operator. The same
formula applies to $Z^R_I$.

The clean separation between coordinate and gauge sectors,
\be
Z_I = Z^R_I + F_I
,
\ee
holds too for higher rank tensors. The operators $Z_I$, $Z^R_I$ and $F_I$ are
all antihermitian. From its definition $Z^R_I$ has the property
\be
   [Z^R_I,V_{\mu_1\mu_2\cdots}] =
   R_{I\mu_1\lambda}V_{\lambda\mu_2\cdots}
   + R_{I\mu_2\lambda}V_{\mu_1\lambda\cdots} + \cdots
   ,
   \label{eq:8.6a}
\ee
where $V$ is a coordinate tensor. In particular for a scalar field
$\phi(x)$
\be
[Z^R_I,\phi] = 0
.
\ee

Now, let $|0\rangle$ be the scalar function that takes the value $1$ for all
$x$. This is a coordinate scalar. The relation $\nabla_\mu|0\rangle=0$ implies
\be
Z^R_{\mu\nu}|0\rangle = 0 = \langle 0|Z^R_{\mu\nu}
\,.
\ee
More generally (using \Eq{B2})
\be
Z^R_I|0\rangle = C_I|0\rangle,
\qquad
\langle 0|Z^R_I = - \langle 0|C_I
,
\ee
with
\bes
C_{\mu\nu} & =0,
\\
C_{\alpha\mu\nu} &= \frac{1}{2}R_{\lambda\mu\nu\alpha\lambda}
,
\\
C_{\alpha_1\cdots\alpha_n\mu\nu}
&=
\frac{1}{2} R_{\lambda \alpha_2\alpha_3\cdots\alpha_n\mu\nu \alpha_1 \lambda}
+
\frac{1}{2} R_{\alpha_1\lambda \alpha_3\cdots\alpha_n\mu\nu \alpha_2 \lambda}
+ \cdots
\\&\quad
+
\frac{1}{2} R_{\alpha_1\cdots\alpha_{n-1}\lambda\mu\nu \alpha_n \lambda}
.
\ees

\subsection{\textsf{Relation with the operator $\RBV_{\mu\nu}$  of
    \cite{Barvinsky:1985an} }}

The operator $Z^R_{\mu\nu}$ coincides with $\RBV_{\mu\nu}$ in
\cite{Barvinsky:1985an}. The higher rank operators
$
(\nabla_{\alpha_1}\cdots \nabla_{\alpha_n}
\RBV_{\mu\nu} )^\rho{}_\sigma
$
are introduced in \cite{Barvinsky:1985an}, with the convention that the
covariant derivative connections do not act on the matrix indices $\rho,\sigma$.
Using the notation
\be
\RBV_{\alpha_1\cdots\alpha_n \mu\nu} \equiv
\nabla_{\alpha_1}\cdots \nabla_{\alpha_n}
\RBV_{\mu\nu} 
\ee
these operators fulfill the recursion
\be
\RBV_{\alpha I} = [\nabla_\alpha,\RBV_I] + R_{I\alpha\lambda} \nabla_\lambda
,
\ee
as well as
\bes
   [\RBV_I,V_{\mu_1\mu_2\cdots}]
   & = [Z^R_I,V_{\mu_1\mu_2\cdots}]
\\&   =
   R_{I\mu_1\lambda}V_{\lambda\mu_2\cdots}
   + R_{I\mu_2\lambda}V_{\mu_1\lambda\cdots} + \cdots
\ees
The operators $Z^R_I$ and $\RBV_I$ are related through
\be
Z^R_I = \RBV_I + C_I
.
\ee
Therefore
\be
\RBV_I |0\rangle = 0, \qquad
\langle 0| \RBV_I = -2 \langle 0|C_I
.
\ee
The operators $\RBV_I$ and $Z^R_I$ have identical commutation properties on
purely multiplicative fields. The $\RBV_I$ are simpler than $Z^R_I$ when
acting on states on the right (since they vanish on coordinate scalars), while
the $Z^R_I$ have the virtue of being antihermitian.

\subsection{\textsf{Derivation of \Eqs{8.17}}}

The first relation in \nec{8.17}, the trace in the coordinate-scalar space,
follows from the fact that $Z^R_I$ coincides with $-C_I$ when acting on
scalars on the left,
\be
\langle\phi|Z^R_I = \langle 0|\phi Z^R_I = \langle 0| ( Z^R_I \phi- [
Z^R_I,\phi] ) = -\langle \phi| C_I
.
\ee

For the second relation, the trace in the coordinate-vector space, consider an operator $\cO^\mu{}_\nu$
acting on the coordinate-vector space, $(\cO V)^\mu=\cO^\mu{}_\nu V^\nu$.
Disregarding the gauge-space sector for simplicity, the trace can be written
as
\be
\tr_1( \cO^\mu{}_\nu ) = \sum_A u^A_\mu \cO^\mu{}_\nu u^\nu_A
\ee
where $u_A^\mu(x)$ is any local basis of vectors at $x$ and $u^A_\mu(x)$ its
dual basis, $u^A_\mu u_B^\mu = \delta^A_B$. When $\cO^\mu{}_\nu$ is purely
multiplicative (i.e., it does not contain $\nabla_\mu$ nor $Z^R_I$), the trace
is simply
\be
\tr_1( \cO^\mu{}_\nu ) = \cO^\mu{}_\nu  u^A_\mu u_A^\nu
= \cO^\mu{}_\nu g^\nu{}_\mu = \cO^\mu{}_\mu
.
\ee

If the operator has a factor $Z^R_I$ on the left, where $I$ is any string of
coordinate indices, without loss of generality we can assume that it has the
form $Z^R_I \cH^{I\mu}{}_\nu$. That is, the row-column indices $\mu\nu$ are not
in $Z_I^R$, and all the indices in $I$ are different. For instance
$Z^R_\alpha{}^{\alpha\mu} X_\nu$ can be rewritten as
$Z^R_{\alpha\beta\lambda} g^{\alpha\beta} g^{\mu\lambda} X_\nu $.
Using now
%
\bes%
\langle u^A_\mu| Z^R_I
&%
= \langle 0| u^A_\mu  Z^R_I
= \langle 0| ( Z^R_I u^A_\mu - [Z^R_I,u^A_\mu] )
\\&%
=
\langle 0| (-C_I u^A_\mu + R_I{}^\lambda{}_\mu u^A_\lambda )
=
\langle u^A_\lambda| (R_I{}^\lambda{}_\mu - g^\lambda{}_\mu C_I)
,
\ees%
it follows that 
%
\bes%
\tr_1( Z^R_I \cO^\mu{}_\nu )
&%
= \sum_A u^A_\mu Z^R_I \cO^\mu{}_\nu u^\nu_A
= \sum_A u^A_\mu (R_I{}^\mu{}_\lambda - g^\mu{}_\lambda C_I)
\cO^\lambda{}_\nu u^\nu_A
\\&%
=  \tr_1\left((R_I{}^\mu{}_\lambda - g^\mu{}_\lambda C_I)  \cO^\lambda{}_\nu \right)
.
\ees%
This proves \Eqs{8.17}. If the operator contains more than one factor $Z^R$ on
the left, the procedure is applied recursively,
%
\bes%
&%
\tr_1( Z^R_I Z^R_J \cO^\mu{}_\nu )
\\&%
=  \tr_1\left((R_I{}^\mu{}_\lambda - g^\mu{}_\lambda C_I)  Z^R_J
\cO^\lambda{}_\nu \right)
\\&%
=
\tr_1\left(Z^R_J (R_I{}^\mu{}_\lambda - g^\mu{}_\lambda C_I)
  \cO^\lambda{}_\nu
-[Z^R_J , R_I{}^\mu{}_\lambda - g^\mu{}_\lambda C_I ]
\cO^\lambda{}_\nu \right)
.
\ees%

\section{\textsf{Canonical form of $\Gamma_S$}
  \label{app:c}}

The following formulas display the expression of the operators in
$\Gamma_{S,n}$, $n=1,2,3,4$ in \Eq{8.1} using the basis of operators in
\nec{8.6} to bring them to the canonical form in \nec{8.4}, by means of the
commutation identities in \nec{8.3}.
These expressions hold in any spacetime dimension.  Eqs. \neq{C1},
\neq{C2}, \neq{C3}, and \neq{C4} correspond to $\Gamma_{L,1}$, $\Gamma_{L,2}$,
$\Gamma_{L,3}$, and $\Gamma_{L,4}$, respectively.

\begin{widetext}%
\bes
\Deltag Y_{\mu\nu } &=
\cO_{19} Y_{\mu\nu }
\\ 
\DeltaM W &= 
\cO_{2} \,W
\\
\frac{1}{2} \DeltaM \{ P_{\mu} ,\nabla_{\mu} \}
&=
(\cO_1)_\mu P_{\mu}
 - \frac{1}{2} \cO_2 \, P_{\mu \mu}
 \label{eq:C1}
 \ees
 
 \bes
 \Deltag Y_{\alpha \mu} \Deltag Y_{\mu \beta}
 &=
\cO_{20}  \, Y_{\alpha \mu} Y_{\mu \beta}
 + O(\nabla^{-5})
 \\ 
 \DeltaM \Phi_{\mu} \Deltag \Phi_{\mu} &=
 \cO_{10}  \, \Phi_{\mu}^2
 + O(\nabla^{-5})
 \\
 \DeltaM \nabla_{\mu} \cM_{\mu \nu} \Deltag \cM_{\nu \alpha} \nabla_{\alpha}
 &= 
 (\cO_3)_{\mu\nu} \,
 \cM_{\mu \alpha} \cM_{\alpha \nu}
 + 2 (\cO_6)_{\mu\nu\alpha} \,
 \cM_{\nu \mu \beta} \cM_{\beta \alpha}
 + 4 (\cO_{12})_{\mu\nu\alpha\beta} \,
 \cM_{\alpha \nu \mu \rho} \cM_{\rho \beta}
 \\&\quad 
 - (\cO_{11})_{\mu\nu} \,
 ( 2 \cM_{\nu \mu \alpha} \cM_{\beta \alpha \beta}
 + \cM_{\alpha \alpha \mu \beta} \cM_{\beta \nu}
 + 2 \cM_{\alpha \nu \mu \beta} \cM_{\beta \alpha}
 + \cM_{\alpha \beta} \cM_{\beta \nu} \cR_{\mu \alpha}
 \\ &\quad\quad
 - 2 F_{\mu \alpha} \cM_{\alpha \beta} \cM_{\beta \nu})
 -  (\cO_{5})_{\mu} \,
 ( \cM_{\mu \nu} \cM_{\alpha \nu \alpha}
 + \cM_{\nu \mu \alpha} \cM_{\alpha \nu})
+ O(\nabla^{-5})
 \\
 \DeltaM \Phi_{\mu} \Deltag \cM_{\mu \nu} \nabla_{\nu} 
 &=
  (\cO_{5})_{\mu} \,
 \Phi_{\nu} \cM_{\nu \mu} 
 + \cO_{10} \,
 (-\Phi_{\mu} \cM_{\nu \mu \nu} - \Phi_{\mu \nu} \cM_{\nu \mu}) 
 + 2 (\cO_{11})_{\mu\nu} \,
 \Phi_{\mu \alpha} \cM_{\alpha \nu}
+ O(\nabla^{-5})
 \\
 \DeltaM \nabla_{\mu} \cM_{\mu \nu} \Deltag \Phi_{\nu} 
 &=
  (\cO_{5})_{\mu} \,
 \cM_{\mu \nu} \Phi_{\nu} 
 + 2 (\cO_{11})_{\mu\nu} \,
 \cM_{\nu \mu \alpha} \Phi_{\alpha}
+ O(\nabla^{-5})
\\
\DeltaM W \DeltaM W
&=
  (\cO_{13}[W])_{\mu\nu} \,
  W
\\ 
\frac{1}{4} \DeltaM \{ P_{\mu}, \nabla_{\mu} \} \DeltaM \{ P_{\nu}, \nabla_{\nu}\}
&=
  (\cO_{4}[P_\mu])_{\mu\nu} \,
P_{\nu}
+\frac{1}{2}  (\cO_{7}[P_{\mu\mu}])_{\nu} \,
P_{\nu}
- (\cO_{16}[P_{\mu},M_{\mu \nu \nu \alpha}])_{\alpha\beta} \,
 P_{\beta}
 \ignore{
   \\&\quad
 - \DeltaM P_{\mu} \DeltaM F_{\mu \nu} M_{\nu \alpha} \DeltaM 
 \nabla_{\alpha} \nabla_{\beta} P_{\beta}
 - \DeltaM P_{\mu} \DeltaM M_{\nu \alpha} F_{\mu \alpha} \DeltaM 
 \nabla_{\nu} \nabla_{\beta} P_{\beta}
 \\&\quad 
+ \DeltaM P_{\mu} \DeltaM M_{\nu \alpha} \cR_{\mu \alpha} \DeltaM 
\nabla_{\nu} \nabla_{\beta} P_{\beta}
}
- (\cO_{9}[P_{\mu},M_{\mu \nu \alpha}])_{\nu\alpha\beta} \,
P_{\beta}
 \\&\quad 
 - (\cO_{16}[P_{\mu},M_{\nu \alpha} R_{\mu \nu \alpha \beta}])_{\beta\rho} \,
 P_{\rho}
 \ignore{
+ \DeltaM P_{\mu} \DeltaM M_{\nu \alpha} R_{\mu \alpha \nu \beta} \DeltaM 
\nabla_{\beta} \nabla_{\rho} P_{\rho}
}
\\&\quad 
+ (\cO_{18}[P_{\mu}, M_{\mu \nu \alpha}, M_{\nu \beta \rho}])_{\beta\rho\alpha\sigma} \,
 P_{\sigma}
 + (\cO_{18}[P_{\mu}, M_{\mu \nu \alpha}, M_{\nu \beta \rho}])_{\alpha\beta\rho\sigma} \,
 P_{\sigma}
\\&\quad 
- \frac{1}{4} \cO_{13}[P_{\mu\mu}] \,
P_{\nu \nu}
- \frac{1}{2} (\cO_{7}[P_{\mu}])_\mu \,
P_{\nu \nu}
+ \frac{1}{2} (\cO_{16}[P_{\mu},M_{\mu \nu \alpha}])_{\nu\alpha} \,
P_{\beta \beta}
 + O(\nabla^{-5})
\\
\frac{1}{2} \DeltaM \{ P_{\mu}, \nabla_{\mu} \} \DeltaM W
&= (\cO_{7}[P_{\mu}])_\mu \,
W
+ \frac{1}{2} \cO_{13}[P_{\mu\mu}] \,
W
- (\cO_{16}[P_{\mu},M_{\mu \nu \alpha}])_{\nu\alpha} \,
W
 + O(\nabla^{-5})
 \label{eq:C2}
 \ees

 \bes
 \DeltaM \nabla_{\mu} \cM_{\mu \nu} \Deltag Y_{\nu \alpha} \Deltag \cM_{\alpha
   \beta} \nabla_{\beta}
 &= (\cO_{11})_{\mu\nu} \,
 \cM_{\mu \alpha} Y_{\alpha \beta} \cM_{\beta \nu} 
 + O(\nabla^{-5})
 \\
 \DeltaM W \DeltaM \nabla_{\mu} \cM_{\mu \nu} \Deltag \cM_{\nu \alpha}
 \nabla_{\alpha} 
 &= (\cO_{14}[W])_{\mu\nu} \,
 \cM_{\mu \alpha} \cM_{\alpha \nu} 
 + O(\nabla^{-5})
 \\
\frac{1}{2} \DeltaM \{ P_{\mu}, \nabla_{\mu} \} \DeltaM \nabla_{\nu} \cM_{\nu 
  \alpha} \Deltag \cM_{\alpha \beta} \nabla_{\beta}
&= 
\frac{1}{2} (\cO_{14}[P_{\mu \mu}])_{\nu\alpha} \,
\cM_{\nu \beta} \cM_{\beta \alpha}
+ (\cO_{8}[P_{\mu}])_{\mu\nu\alpha} \,
\cM_{\nu \beta} \cM_{\beta \alpha}
 \\&\quad 
 + 2 (\cO_{15}[P_{\mu}])_{\mu\nu\alpha\beta} \,
 \cM_{\alpha \nu \rho} \cM_{\rho \beta}
 \\&\quad
 - (\cO_{17}[P_{\mu}, M_{\mu \nu \alpha} ])_{\nu\alpha\beta\rho} \,
 \cM_{\beta \sigma} \cM_{\sigma \rho}
 \\&\quad 
 - (\cO_{14}[P_{\mu} ])_{\mu\nu} \,
 ( \cM_{\nu \alpha} \cM_{\beta \alpha \beta}
 + \cM_{\alpha \nu \beta} \cM_{\beta \alpha})
 + O(\nabla^{-5})
 \\ 
 \frac{1}{2} \DeltaM \{ P_{\mu}, \nabla_{\mu} \} \DeltaM \nabla_{\nu} 
 \cM_{\nu \alpha} \Deltag \Phi_{\alpha}
 &= 
(\cO_{14}[P_{\mu} ])_{\mu\nu} \,
 \cM_{\nu \alpha} \Phi_{\alpha} 
 + O(\nabla^{-5})
 \\
\frac{1}{2} \DeltaM \{ P_{\mu}, \nabla_{\mu} \} \DeltaM \Phi_{\nu} \Deltag 
 \cM_{\nu \alpha} \nabla_{\alpha}
 &=
(\cO_{14}[P_{\mu} ])_{\mu\nu} \,
 \Phi_{\alpha} \cM_{\alpha \nu} 
 + O(\nabla^{-5})
 \\
 \frac{1}{4} \DeltaM W \DeltaM \{ P_{\mu}, \nabla_{\mu} \} \DeltaM \{ P_{\nu}, 
 \nabla_{\nu} \}
 &= 
(\cO_{16}[W , P_{\mu} ])_{\mu\nu} \,
 P_{\nu}
 + O(\nabla^{-5})
 \\
\frac{1}{8} \DeltaM \{ P_{\mu}, \nabla_{\mu} \} \DeltaM \{ P_{\nu},
\nabla_{\nu} \} \DeltaM \{ P_{\alpha}, \nabla_{\alpha} \}
&=
(\cO_{16}[P_{\mu}, P_{\mu\nu} ])_{\nu\alpha} \,
P_{\alpha} 
+ \frac{1}{2} (\cO_{16}[P_{\mu}, P_{\nu\nu} ])_{\mu\alpha} \,
P_{\alpha}
 \\&\quad
 + \frac{1}{2} (\cO_{16}[P_{\mu\mu}, P_{\nu} ])_{\nu\alpha} \,
 P_{\alpha}
 + (\cO_{9}[P_{\mu}, P_{\nu} ])_{\mu\nu\alpha} \,
 P_{\alpha}
 \\&\quad 
 - (\cO_{18}[P_{\mu}, M_{\mu \nu \alpha}, P_{\beta}])_{\nu\alpha\beta\rho} \,
 P_{\rho}
 \\&\quad 
 - (\cO_{18}[P_{\mu}, P_\nu , M_{\nu \alpha\beta} ])_{\alpha\beta\nu\rho} \,
 P_{\rho}
 \\&\quad 
 - (\cO_{18}[P_{\mu}, P_\nu , M_{\nu \alpha\beta} ])_{\mu\alpha\beta\rho} \,
 P_{\rho}
 \\&\quad 
 - \frac{1}{2} (\cO_{16}[P_{\mu}, P_\nu ])_{\mu\nu} \,
 P_{\alpha \alpha}
 + O(\nabla^{-5})
 \label{eq:C3}
 \ees

 \bes
 \DeltaM \nabla_{\mu} \cM_{\mu \nu} \Deltag \cM_{\nu \alpha} 
 \nabla_{\alpha} \DeltaM \nabla_{\beta} \cM_{\beta \rho} \Deltag 
 \cM_{\rho \sigma} \nabla_{\sigma} 
 &=
 (\cO_{15}[\cM_{\mu \nu} \cM_{\nu \alpha} ])_{\mu\alpha\beta\rho} \,
 \cM_{\beta \sigma} \cM_{\sigma \rho}
 + O(\nabla^{-5})
 \\
\frac{1}{4} \DeltaM \{ P_{\mu}, \nabla_{\mu} \} 
 \DeltaM \{ P_{\nu}, \nabla_{\nu} \} 
 \DeltaM \nabla_{\alpha} \cM_{\alpha \beta} \Deltag \cM_{\beta \rho} 
 \nabla_{\rho}
 &=
 (\cO_{17}[P_{\mu},  P_{\nu} ])_{\mu\nu\alpha\beta} \,
 \cM_{\alpha \rho} \cM_{\rho \beta}
 + O(\nabla^{-5})
 \\ 
 \frac{1}{16}\DeltaM \{ P_{\mu}, \nabla_{\mu} \} \DeltaM \{ P_{\nu}, 
 \nabla_{\nu} \} \DeltaM \{ P_{\alpha}, 
 \nabla_{\alpha} \} \DeltaM \{ P_{\beta}, \nabla_{\beta} \}
 &= 
 (\cO_{18}[P_{\mu},  P_{\nu}, P_{\alpha}])_{\mu\nu\alpha\beta} \,
 P_{\beta} 
 + O(\nabla^{-5})
 \label{eq:C4}
\ees
\end{widetext}%

\section{\textsf{Perturbative expansions}
  \label{app:d}}

For completeness, we give here the expressions for $\Gamma_{L,0}^\div$ and
$\Gamma_S^\div$ in $d=2$ (Eqs. \nec{10.18b} and \nec{8.36b} respectively) to
second order in an expansion in powers of $H^{\mu\nu}$, where
\be M^{\mu\nu}= m^2 g^{\mu\nu}+H^{\mu\nu} \ee
and to all orders in the other fields. $m^2$ is a constant c-number. In $d=2$
the combination $\Gamma^\div_{L,1}+\Gamma_\gh^\div$ cancels, so there are no
further contributions to $\Gamma_1^\div$.

The result is obtained by expanding Eqs. \nec{10.18b} and \nec{8.36b} and
carrying out the angular average, and integration over $z$ in the case of
$\Gamma_{L,0}^\div$, as such evaluations can be done explicitly when $M^{\mu\nu}= m^2 g^{\mu\nu}$.
\begin{widetext}%
\bes
\Gamma_{L,0}^\div &=
\frac{1}{4\pi\epsilon} \Big\langle
\frac{1}{12} \bR 
+ \frac{1}{m^4} \Big(
-\frac{1}{12} F_{\mu \nu }\,H_{\mu \alpha }\,H_{\nu \alpha }
-\frac{1}{96} H_{\mu \mu }\,H_{\nu \nu }\,\bR
+\frac{1}{48} H_{\mu \nu }\,H_{\mu \nu }\,\bR
\\&\quad
-\frac{1}{24} H_{\mu \mu \nu }\,H_{\nu \alpha \alpha}
-\frac{1}{96} H_{\mu \nu \nu }\,H_{\mu \alpha \alpha }
-\frac{1}{48} H_{\mu \nu \alpha }\,H_{\mu \nu \alpha }
+\frac{1}{24} H_{\mu \nu \alpha }\,H_{\nu \mu \alpha }
\Big)
+ O(H^3)
\Big\rangle_{x,g}
.
\ees
\newcommand{\Wt}{\tilde{W}}
\bes
\Gamma_S^\div &=
\frac{1}{4\pi\epsilon} \Big\langle
- \frac{1}{2} m^2
- \frac{1}{2} \bR
- \frac{1}{4} H_{\mu \mu }
\\& \quad
+ \frac{1}{m^2} \Big(
- \frac{1}{2}\Wt - \frac{1}{4} Q_{\mu \nu }\,Q_{\mu \nu }
- \frac{1}{16} H_{\mu \mu }\,H_{\nu \nu }
+ \frac{1}{8} H_{\mu \nu }\,H_{\mu \nu }
     \Big)
\\& \quad
+ \frac{1}{m^4} \Big(
- \frac{1}{8} P_{\mu }\,P_{\mu }
+ \frac{1}{4} \Wt\,H_{\mu \mu }
+ \frac{1}{16} Q_{\mu \nu }\,Q_{\mu \nu }\,H_{\alpha \alpha }
+ \frac{1}{8} Q_{\mu \nu }\,Q_{\mu \alpha }\,H_{\nu \alpha }
+ \frac{1}{4} Q_{\mu \nu }\,H_{\mu \alpha }\,H_{\nu \alpha } 
   \Big)
\\& \quad
+ \frac{1}{m^6} \Big(
 \frac{1}{16} P_{\mu }\,P_{\mu }\,H_{\nu \nu }
+ \frac{1}{8} P_{\mu }\,P_{\nu }\,H_{\mu \nu }
- \frac{1}{24} P_{\mu }\,H_{\mu \nu }\,H_{\nu \alpha \alpha }
+ \frac{1}{24} P_{\mu }\,H_{\mu \nu }\,H_{\alpha \nu \alpha }
+ \frac{1}{96} P_{\mu }\,H_{\nu \nu }\,H_{\mu \alpha \alpha }
\\& \quad\quad
+ \frac{1}{48} P_{\mu }\,H_{\nu \nu }\,H_{\alpha \mu \alpha }
+ \frac{1}{48} P_{\mu }\,H_{\nu \alpha }\,H_{\mu \nu \alpha }
- \frac{1}{12} P_{\mu }\,H_{\nu \alpha }\,H_{\nu \mu \alpha }
-\frac{1}{96} P_{\mu }\,H_{\mu \nu \nu }\,H_{\alpha \alpha }
-\frac{1}{48} P_{\mu }\,H_{\mu \nu \alpha }\,H_{\nu \alpha }
\\& \quad\quad
-\frac{1}{48} P_{\mu }\,H_{\nu \mu \nu }\,H_{\alpha \alpha }
+\frac{1}{12} P_{\mu }\,H_{\nu \mu \alpha }\,H_{\nu \alpha }
-\frac{1}{24} P_{\mu }\,H_{\nu \nu  \alpha }\,H_{\mu \alpha }
+\frac{1}{24} P_{\mu }\,H_{\nu \alpha \alpha }\,H_{\mu \nu }
-\frac{1}{16} \Wt\,H_{\mu \mu }\,H_{\nu \nu }
\\& \quad\quad
-\frac{1}{8} \Wt\,H_{\mu \nu }\,H_{\mu \nu }
-\frac{1}{96} Q_{\mu \nu }\,Q_{\mu \nu }\,H_{\alpha \alpha }\,H_{\beta \beta }
   -\frac{1}{48} Q_{\mu \nu }\,Q_{\mu \nu }\,H_{\alpha \beta }\,H_{\alpha\beta}
   -\frac{1}{48} Q_{\mu \nu }\,Q_{\mu \alpha }\,H_{\nu \alpha }\,H_{\beta\beta}
\\& \quad\quad
   -\frac{1}{24} Q_{\mu \nu }\,Q_{\mu\alpha }\,H_{\nu \beta }\,H_{\alpha \beta }
   -\frac{1}{24} Q_{\mu \nu }\,Q_{\mu \alpha }\,H_{\alpha \beta }\,H_{\nu \beta }
   -\frac{1}{48} Q_{\mu \nu }\,Q_{\mu \alpha }\,H_{\beta \beta }\,H_{\nu \alpha }
   \Big)
\\& \quad
+\frac{1}{m^8} \Big(
  -\frac{1}{96} P_{\mu }\,P_{\mu }\,H_{\nu \nu }\,H_{\alpha \alpha }
  -\frac{1}{48} P_{\mu }\,P_{\mu}\,H_{\nu \alpha }\,H_{\nu \alpha }
  -\frac{1}{48} P_{\mu }\,P_{\nu}\,H_{\mu \nu }\,H_{\alpha \alpha}
  -\frac{1}{24} P_{\mu }\,P_{\nu }\,H_{\mu \alpha }\,H_{\nu \alpha}
  -\frac{1}{24} P_{\mu }\,P_{\nu }\,H_{\nu \alpha }\,H_{\mu \alpha}
  \\& \quad\quad
  -\frac{1}{48} P_{\mu }\,P_{\nu }\,H_{\alpha \alpha }\,H_{\mu \nu}
  -\frac{1}{48} P_{\mu }\,H_{\mu \nu }\,P_{\nu }\,H_{\alpha\alpha}
  -\frac{1}{48} P_{\mu }\,H_{\mu \nu }\,P_{\alpha }\,H_{\nu \alpha}
  -\frac{1}{192} P_{\mu }\,H_{\nu \nu }\,P_{\mu }\,H_{\alpha \alpha}
  \\& \quad\quad
  -\frac{1}{96} P_{\mu }\,H_{\nu \alpha }\,P_{\mu }\,H_{\nu \alpha}
  -\frac{1}{48} P_{\mu }\,H_{\nu \alpha}\,P_{\nu }\,H_{\mu \alpha}
  \Big)
  + O(H^3)
  \Big\rangle_{x,g}
  .
  \ees
  In this formula $\Wt \equiv  W-\frac{1}{2}P_{\mu\mu}$.
\end{widetext}%

The term $\bR$ in $\Gamma_{L,0}^\div$ is the Gauss-Bonnet invariant in two
dimensions, complying by itself with the transverse and longitudinal symmetry
invariance discussed in Sec. \ref{sec:10.D}. In the two-dimensional case
these symmetries do not allow terms with one $H^{\mu\nu}$ nor of order $1/m^2$
in $\Gamma_{L,0}^\div$.

\section{\textsf{Method of non-covariant symbols
    \label{app:e}}}

\subsection{\textsf{Covariant vs non-covariant method of symbols}
}

The method of non-covariant symbols \cite{Nepomechie:1984wt,Salcedo:1994qy}
allows to obtain diagonal matrix elements of pseudo-differential operators and
it can be extended to curved spacetime \cite{Salcedo:2006pv}.

The difference between the covariant and non-covariant versions can be
elucidated already in the case of flat spacetime. Let $\hat{f}=f(D,X)$ be a
pseudo-differential operator constructed out of the gauge covariant derivative
$D_\mu$ and one (or more) non-Abelian fields $X(x)$ (a purely multiplicative
operator). Then
\bes
\esp{x|\hat{f}|x} &= 
\int\frac{d^dp}{(2\pi)^d}\esp{x|\hat{f}|p\rangle\langle p|x}
\\
&= 
\int\frac{d^dp}{(2\pi)^d} e^{-xp}\esp{x|\hat{f}|p}
\\
&= 
\int\frac{d^dp}{(2\pi)^d} \esp{x|e^{-xp}\hat{f}e^{xp}|0}
\\
&= 
\int\frac{d^dp}{(2\pi)^d} \esp{x|f(D+p,X)|0}
,
\ees
where $p_\mu$ is imaginary and $|0\rangle$ is the state $\esp{x|0}=1$. The quantity
$\esp{x|f(D+p,X)|0}$ is the (non-covariant) symbol of $\hat{f}$.

After momentum integration, the result is gauge covariant in the sense that $D_\mu$ will only appear in
the form of a commutator $[D_\mu,~]$. This follows from the fact that under
the shift $D_\mu\to D_\mu+a_\mu$, where $a_\mu$ is an arbitrary constant
imaginary c-number, the dependence on $a_\mu$ cancels upon momentum
integration (since $a_\mu$ can be compensated by a corresponding shift in
$p_\mu$). The virtue of the covariant symbol
\be
\overline{f}=  e^{-D\partial}f(D+p,X)e^{D\partial}
\ee
is that it is multiplicative (with respect to $x$) and is
already covariant without momentum integration \cite{Salcedo:2006pv}.

To illustrate the point let us apply the method of non-covariant symbols to
$\hat{f}= (D_\mu^2-X)^{-1}$,
\be
\esp{x|\hat{f}|x} = \Esp{\frac{1}{(p_\mu+D_\mu)^2-X}}_p
\ee
Defining $N=1/(p_\mu^2-X)$ and expanding in powers of $D_\mu$,
\bes
\esp{x|\hat{f}|x} &=
\Big\langle
  N -N(2p_\mu D_\mu +D_\mu^2)N
  \\&\quad
  +
  N(2p_\mu D_\mu)N(p_\nu D_\nu)N + O(D^3)
\Big\rangle_p
\label{eq:E4}
\ees
Instead of doing the momentum integration one can add terms which are
identically zero by integration by parts in momentum space to bring the
expression to a covariant form. For instance the first order
term\mfootnote{Actually this term vanishes by parity but nevertheless it serves to
  illustrate the point.}
\be
\esp{x|\hat{f}|x}^{(1)} = \Esp{ -2p_\mu N D_\mu N }_p
\ee
can be supplemented with
\be
0= \Esp{ \partial^\mu (- D_\mu N) }_p
= \Esp{ 2 p_\mu D_\mu N^2 }_p,
\ee
yielding a manifestly gauge covariant result
\be
\esp{x|\hat{f}|x}^{(1)} = \Esp{ 2p_\mu [D_\mu, N] N }_p
\label{eq:E7}
.
\ee
This procedure can be carried out systematically. Rewriting \Eq{E4} as
\bes
& \esp{x|\hat{f}|x} = \Esp{T_0 + T_1 + T_2 + \cdots }_p
\\ &
T_0 = N, \quad
T_1= -2N \,pD\, N,
\\ &
T_2= -N\, D D\, N + 4N \, p D \, N \,pD \, N,
\ees
the systematic integration by parts suggested by the method of covariant
symbols can be implemented as\mfootnote{Equivalently, one can put instead a
  factor $e^{pD}$ at the right and move $\partial^\mu$ to left.}
\bes
e^{-D\partial}\sum_n T_n |0\rangle &=
\sum_n \tilde{T}_n |0\rangle
,
\\
\tilde{T}_n &= \sum_{j=0}^n \frac{(-1)^j}{j!} (D\partial)^j T_{n-j}
.
\label{eq:E9}
\ees
and now
\be
\esp{x|\hat{f}|x} = \Esp{\tilde{T}_0 + \tilde{T}_1 + \tilde{T}_2 + \cdots }_p
.
\label{eq:E10}
\ee
In this way
\be
\tilde{T}_0=T_0 = N
\ee
is already covariant,
\be
\tilde{T}_1 = T_1 -D\partial\, T_0 =
-2N\,pD\,N-D\partial\, N
=
2[pD,N]N
\ee
is the term obtained previously in \Eq{E7}. For the second order
\bes
\tilde{T}_2 &= T_2-D\partial\, T_1 + \frac{1}{2} (D\partial)^2T_0
\\&=
-N\,DD\,N+4N\,pD\,N\,pD\,N
-D\partial(-2N\,pD\,N)+
\frac{1}{2}(D\partial)^2N
\ees
Carrying out the momentum derivatives and moving the $D_\mu$ to the left
yields a manifestly covariant form
\be
\tilde{T}_2 =
-N_{\mu\mu} N + 4p_\mu p_\nu (N_\mu N_\nu + N_{\mu\nu}N)N
,
\ee
where
\be
N_\mu = [\nabla_\mu,N], \qquad
N_{\mu\nu} = [ \nabla_\mu , N_\nu ]
.
\ee

\subsection{\textsf{Method of non-covariant symbols in curved spacetime}
}

The results presented in this work, obtained using the method of covariant
symbols, have been reproduced also using the method of non-covariant
symbols. For the latter, the technique in Eqs. \nec{E9} and \nec{E10} is
applied, which produces covariant expressions without requiring further
integration by parts in momentum space. This holds too in curved spacetime
using the covariant derivative $\nabla_\mu$ which includes all connections,
and $\frac{1}{2}\{\nabla_\mu,\partial^\mu\}$ instead of $D_\mu\partial^\mu$,
in the exponential.

In this approach typically the $\partial^\mu$ (generated in \Eq{E9}) are
removed by moving them to the right. Then the free $\nabla_\mu$ are moved
together, say to the right, to form covariant combinations
$Z_{\mu\nu}=[\nabla_\mu,\nabla_\nu]$, etc. Therefore one has to specify how
$\nabla_\mu$ commutes with $p_\nu$ and $\partial^\nu$, as these commutators
do not vanish (consistently with
$[Z_{\mu\nu},p_\alpha] = R_{\mu\nu\alpha\lambda}p_\lambda$).

To this end let us introduce a set of coordinates $\xi^A(x)$ (technically $d$
coordinate scalars). Eventually, after all free covariant derivative operators
have been removed, these will be the Riemann coordinates at $x_0$ (the point
at which the diagonal matrix element is being computed). And let the vector
fields $t^A_\mu(x)$ and $t^\mu_A(x)$ be defined through the
relations
\be
t^A_\mu = [\nabla_\mu ,\xi^A], \qquad
t_\mu^A \, t_B^\mu = \delta^A_B
.
\ee
Note that the $d$ vector fields $t_\mu^A(x)$ do not define a tetrad. In terms
of these
\be
p_\mu = t_\mu^A p_A,\qquad
\partial^\mu = t^\mu_A \partial^A,\qquad
\partial^A\equiv \partial/\partial p_A
\ee
where $p_A$ are (imaginary) constant c-number parameters, hence
$[\nabla_\mu,p_A]=[\nabla_\mu,\partial^A]=0$. This allows to write
\bes
    [\nabla_\mu,p_\nu]  &= t^A_{\mu\nu}p_A = t^\lambda_A t^A_{\mu\nu}p_\lambda
    \equiv -G_\mu{}^\lambda{}_\nu p_\lambda
    \\
      [\nabla_\mu,\partial^\nu]  &=  G_\mu{}^\nu{}_\lambda \partial^\lambda
      ,
\ees
using $[\nabla_\mu, t^\nu_A]=-t^\nu_B \, t^B_{\mu\alpha} \, t^\alpha_A$ in the
second relation.  Note that $t_{\mu\nu}^A=t_{\nu\mu}^A$, hence
$G_{\mu\lambda\nu} = G_{\nu\lambda\mu}$. Successive derivatives require
derivatives of $G_{\mu\nu\alpha}(x)$ which in turn follow from those of
$t_\mu^A(x)$. Here it enters the Riemann coordinate condition at $x_0$, which
requires~\cite{Gusynin:1990bu}
\begin{itemize}
\item[$i)$] $t_\mu^A = \delta^A_\mu$ at $x_0$, and
\item[$ii)$] the vanishing of the completely symmetry component of
  $t^A_{\mu_1\cdots\mu_n}$ at $x_0$ for $n\ge 2$.
\end{itemize}
From these conditions, using the Jacobi identity \nec{8.19}, it follows
\bes
t^A_{\mu\nu}\big|_{x_0} &= 0
\\
t^A_{\alpha\mu\nu}\big|_{x_0} &= \frac{1}{3} \left(
R_{\alpha\mu\nu}{}^\lambda
+
R_{\alpha\nu\mu}{}^\lambda \right) t^A_\lambda
\\
t^A_{\alpha\beta\mu\nu}\big|_{x_0} &=  \Big(
\frac{1}{4} R_{\alpha\beta\mu\nu}{}^\lambda
+\frac{1}{4} R_{\alpha\beta\nu\mu}{}^\lambda
+\frac{1}{6} R_{\beta\alpha\mu\nu}{}^\lambda
+\frac{1}{6} R_{\beta\alpha\nu\mu}{}^\lambda
\\&\qquad
+\frac{1}{12} R_{\mu\alpha\nu\beta}{}^\lambda
+\frac{1}{12} R_{\nu\alpha\mu\beta}{}^\lambda
\Big) t^A_\lambda
.
\ees
Hence,
\bes
G_{\mu\lambda\nu}(x_0) &= 0
\\
G_{\alpha\mu\lambda\nu}(x_0) &= -\delta^\lambda_A \,t^A_{\alpha\mu\nu}(x_0)
\\
G_{\alpha\beta\mu\lambda\nu}(x_0) &=
-\delta^\lambda_A \,t^A_{\alpha\beta\mu\nu}(x_0)
.
\ees
It is important to remark that the conditions at $x=x_0$ can only be imposed
after all free $\nabla_\mu$ have been removed, and also that the $p_\mu$'s in
$N_g$ and $N_M$ have to be differentiated too, $p_\mu^2$ cannot treated as a
constant during this calculation. On the other hand, as proven in App. C of
\cite{Garcia-Recio:2019iia}, in the final expression (i.e., after $x$ is set
to $x_0$) one can freely integrate by parts neglecting derivatives of $p_\mu$.

\section{\textsf{ Sample calculation using the method of covariant
    symbols }
  \label{app:f}}

As the method of covariant symbols is not well known we will present here the
calculation of one of the universal functional traces of 
\cite{Barvinsky:1985an} using this method. Concretely we consider
\be
\esp{x | \nabla_\mu \Delta |x}^\div_{d=4}
\ee
We want the UV divergent part in $4+2\epsilon$ dimensions. Therefore we need
the terms of order $1/p^4$ of the covariant symbol of $\nabla_\mu
\Delta$. Since this operator is of order $1/p$, and
the expansion of the covariant symbols is in powers of $\nabla/p$ it
will necessary to go to order  $(\nabla/p)^3$. That is
\be
(\overline{\nabla_\mu \Delta})_{-4} =
(\overline{\nabla}_\mu)_{1} (\overline{\Delta})_{-5}
+ (\overline{\nabla}_\mu)_{-2} (\overline{\Delta})_{-2}
,
\label{eq:F2}
\ee
where the subindex indicates the number of $p_\mu$ minus the number of
$\partial^\mu$ in the term. Note that $(\overline{\nabla}_\mu)_{0}=0$, as can
be seen in \Eq{8.13a}, and also \,$(\overline{\Delta})_{-3}=0$  \,due to
\,$(\overline{\nabla^2})_{1}=0$.

Eqs. (B1) of \cite{Garcia-Recio:2019iia} show the expressions of
$(\overline{\nabla}_\mu)_{n}$ for $n=1,\ldots,-3$ and of
$(\overline{\nabla^2}_\mu)_{n}$ for $n=2,\ldots,-2$ (i.e., to four covariant
derivatives). The expressions of $(\overline{\Delta}_\mu)_{n}$ are not
tabulated there but they need to be computed only once, using
\bes
\overline{\Delta} &= (\overline{\nabla^2})^{-1}
\\
&=
\left(
(\overline{\nabla^2})_{2}
+
(\overline{\nabla^2})_{0}
+
(\overline{\nabla^2})_{-1}
+O(p^{-2})
\right)^{-1}
.
\ees
This gives
\be
(\overline{\Delta})_{-2} = (p_\mu p_\nu g^{\mu\nu})^{-1} = -N_g
\ee
(with $N_g$ already defined in \nec{8.14a}) and
\be
(\overline{\Delta})_{-5} = -N_g \, (\overline{\nabla^2})_{-1} \, N_g
\,.
\ee

The formulas $(\overline{\nabla}_\mu)_{1} = p_\mu$ and
$(\overline{\Delta})_{-2} = -N_g$, as well as the tabulated values of
$(\overline{\nabla}_\mu)_{-2}$ and $(\overline{\nabla^2})_{-1}$, are inserted
in \Eq{F2} and this results in an expression containing $\partial_\mu$. These
derivatives are removed by using
\be
[ \partial^\mu , N_g]  = 2p^\mu N_g^2
\ee
and \nec{12.4}. Also this very equation is used to move all
$Z_{\mu_1\cdots\mu_n}$ to the right. This procedure yields
\bes
\esp{x | \nabla_\mu \Delta |x}^\div_{d=4}
&=
\Big\langle
-\frac{2}{3} N_g^2    Z_{\nu   \nu   \mu }
- \frac{2}{3} N_g^3 p_{\mu }  p_{\nu }  Z_{\alpha \alpha \nu }
\\&
-  \frac{8}{3} N_g^3  p_{\nu }  p_{\alpha }  Z_{\nu \alpha \mu }
+   \frac{1}{2} N_g^3   p_{\mu }  p_{\nu }  \bR_{\nu }
\\&
+  N_g^3   p_{\nu }  p_{\alpha } \cR_{\mu   \nu   \alpha }
-  2 N_g^3   p_{\nu }  p_{\alpha }  \cR_{\nu \alpha \mu }
\Big\rangle_p
\ees
The integral over $p_\mu$ is immediate and gives
\be
\esp{x | \nabla_\mu \Delta |x}^\div_{d=4}
  =
- \frac{1}{(4\pi)^2 \epsilon}
\left(
\frac{1}{6} Z_{\alpha\alpha\mu} - \frac{1}{8} \bR_\mu
\right)
,
\ee
using $\cR_{\nu\nu\mu}=\frac{1}{2}\bR_\mu$.

This result can be expressed in terms of the operator $\RBV_I$. Using the
relations
$Z_I=F_I + Z^R_I$ and $Z^R_I = \RBV_I + C_I$, and also
$C_{\alpha\alpha\mu} =
\frac{1}{4}\bR_\mu$, from \Eq{B8}, one obtains
\be
\esp{x | \nabla_\mu \Delta |x}^\div_{d=4}
  =
- \frac{1}{(4\pi)^2 \epsilon}
\left(
  \frac{1}{6} F_{\alpha\alpha\mu} + \frac{1}{6} \RBV_{\alpha\alpha\mu}
  - \frac{1}{12} \bR_\mu
\right)
,
\ee
a result in agreement with Eq. (4.54) of \cite{Barvinsky:1985an}. Other
universal functional traces are also reproduced with the method of covariant
symbols.


\end{document}